\pdfoutput=1 
\documentclass[a4paper,fleqn,usenatbib]{mnras}

\usepackage{newtxtext,newtxmath}

\usepackage[T1]{fontenc}
\usepackage{ae,aecompl}
\usepackage{lscape}

\usepackage{graphicx}	
\usepackage{amsmath}	
\usepackage{amssymb}	
\usepackage{subfigure}
\usepackage{float}
\usepackage{multirow}
\hypersetup{citecolor=magenta}

\title[ICL: a luminous DM tracer]{Intracluster light: a luminous tracer for dark matter in clusters of galaxies.}

\author[M. Montes \& I. Trujillo]{
Mireia Montes,$^{1,2}$\thanks{E-mail: mireia.montes.quiles@gmail.com (MM)}
and Ignacio Trujillo,$^{3,4}$
\\
$^{1}$School of Physics, University of New South Wales, Sydney, NSW 2052, Australia\\
$^{2}$Department of Astronomy, Yale University, New Haven, CT, 06511, USA\\
$^{3}$Instituto de Astrof\'isica de Canarias, c/ V\'ia L\'actea s/n, E-38205, La Laguna, Tenerife, Spain\\
$^{4}$Departamento de Astrof\'isica, Universidad de La Laguna, E-38205 La Laguna, Tenerife, Spain}

\date{Accepted XXX. Received YYY; in original form ZZZ}

\pubyear{2018}

\begin{document}
\label{firstpage}
\pagerange{\pageref{firstpage}--\pageref{lastpage}}
\maketitle

\begin{abstract}

The bulk of stars in galaxy clusters are confined within their constituent galaxies. Those stars do not trace the extended distribution of dark matter well as they are located in the central regions of the cluster's dark matter sub-halos. A small fraction of stars is expected, however, to follow the global dark matter shape of the cluster. These are the stars whose extended spatial distribution results from the merging activity of galaxies and form the intracluster light (ICL). In this work, we compare the bi-dimensional distribution of dark matter in massive galaxy clusters (as traced by gravitational lensing models) with the distribution of the ICL. To do that, we use the superb data from the Hubble Frontier Fields Initiative. Using the Modified Hausdorff distance (MHD) as a way of quantifying the similarities between the mass and ICL distributions, we find an excellent agreement (MHD$\sim$25 kpc) between the two components. This result shows that the ICL exquisitely follows the global dark matter distribution, providing an accurate luminous tracer of dark matter. This finding opens up the possibility of exploring the distribution of dark matter in galaxy clusters in detail using only deep imaging observations.
\end{abstract}





\section{Introduction}

\label{sec:intro}

One of the first evidences of non baryonic dark matter (DM) was found in galaxy clusters \citep[][]{Zwicky1933,Zwicky1937}. This collisionless DM makes up most of the mass in the Universe \citep[][]{PlanckCollaboration2016a}, does not radiate and only interacts gravitationally with visible matter providing only indirect ways of detecting it. Since DM is an essential constituent of cosmological theories, its distribution within galaxy clusters would help us to distinguish not only among different scenarios on the nature of DM itself \citep[i.e. warm vs cold particles; ][]{Primack1984} but also on different alternatives to DM as modified gravity \citep[see e.g.][]{Milgrom2002,McGaugh2015}.

In this sense, gravitational lensing has proven to be an invaluable tool to study the DM distribution within clusters of galaxies \citep[see for a review][]{Kneib2011, Hoekstra2013}. Gravitational lensing helps us to understand structure formation, probes the nature of DM and fully captures the interplay between baryons and DM \citep[e.g.][]{Clowe2004, Markevitch2004}. In particular, gravitational lensing offers a unique and powerful probe of the substructure of the DM in galaxy clusters, independently of the dynamical state of the object producing the lensing. Despite these great advantages, having access to the detailed mass distributions of the galaxy clusters through gravitational lensing is very demanding observationally. An accurate gravitational lensing reconstruction requires not only deep imaging to identify multiply-lensed images but also long spectroscopic campaigns to confirm the redshift of these images. In this sense, finding an alternative observational proxy to trace the DM distribution in galaxies would be ideal.

X-rays might appear as a way to address this problem. Galaxy clusters emit in X-rays due to thermal bremsstrahlung produced in the highly ionised gas bound by the gravitational potential well of the cluster. This powerful emission is directly linked to the total gravitating mass they contain, such that they can be efficiently used as tracers of the cosmic distribution of mass within a considerable fraction of the observable Universe \citep[e.g.][]{Borgani2001}. While in more relaxed clusters it might be true that the X-rays follow the DM distribution, in galaxy clusters that are undergoing a merger process, the gas (dissipative) experiences ram pressure and is slowed, creating an offset between the DM and the X-ray emission. A well-known example of this is the Bullet Cluster \citep[e.g.][]{Clowe2004, Markevitch2004}. For this reason, it is worth exploring an alternative luminous tracer for the detailed distribution of DM that improves the one provided by the X-ray maps, particularly in those galaxy clusters that are not relaxed (i.e. the majority of cases).

In the current cosmological paradigm ($\Lambda$CDM), clusters of galaxies are assembled hierarchically by the accretion of galaxies or small galaxy groups. Observationally, one of the most revealing signatures of this assembly is the ICL (see \citealt{Mihos2016} for a review). This diffuse light is composed of a substantial fraction of stars, between $5-20$\% of the total amount of stars in the cluster, \citep[e.g.][]{Krick2007,Burke2015,Montes2018,Jimenez-Teja2018}. The ICL forms primarily by the interaction and merging of satellite galaxies during the assembly of the cluster \citep[e.g.][]{Gregg1998, Mihos2005, MT14, Montes2018}. The physical scales of the ICL, several hundreds of kpc, are similar to those of the DM distribution \citep[e.g.][]{Dubinski1998}, so it is reasonable to expect that this component will help us trace the global gravitational potential of its host cluster. In fact, \citet[][]{Pillepich2014, Pillepich2018} used the Illustris and IllustrisTNG suites of simulations to explore stellar halos in systems encompassing a wide range of masses. In their analysis, they found a correlation between the logarithmic slope of the stellar density profile at large radius\footnote{The radius used in \citet{Pillepich2018} to fit this slope of the density profile ranges between $30 kpc < R < R_{200c}$, with $R_{200c}$ being the 3D virial radius.} (i.e. in the region dominated by the stellar halo) and the total mass of the halo. Furthermore, they claimed that this slope can be as shallow as the underlying DM slope for masses as large as $M_h = 10^{14-15} M_{\odot}$. That is, both stellar and DM haloes have similar shapes. This is a direct consequence of the hierarchical assembly of galaxy cluster. More massive halos form later which means that they are less concentrated \citep[e.g.][]{Navarro1996, Navarro1997, Gao2004}. Consequently, if the host halos are less concentrated a satellite infalling into the cluster gravitational potential deposit its stripped stars at large radius \citep[see also][]{Cooper2015}. The \citet[][]{Pillepich2014, Pillepich2018} prediction has been recently confirmed in \citet[][hereafter MT18]{Montes2018}. In such work, MT18 explore the correlation between the slope of the stellar density profile of the ICL and the total mass of the halo (M$_{200}$) in the Hubble Frontier Fields (HFF) clusters. They found that the slopes of the ICL of the HFF clusters follow the extrapolation of the theoretical expectations in \citet[][]{Pillepich2018}.



Motivated by the above result, we decided to take a step forward and explore whether the ICL can be used to trace in detail the DM distribution (including substructures) in galaxy clusters. In the present work, we use the accurate mass maps built using gravitational lensing and compare those with the ICL bi-dimensional distribution. In order to do that, we use the exquisite data provided by the HFF Initiative \citep[][]{Lotz2017}. The HFF Initiative appears as the perfect dataset for exploring this as it not only provides the deepest images of six galaxy clusters ever observed with the \emph{Hubble Space Telescope} (HST), but it also provides accurate gravitational lensing models of the clusters. The hundreds of multiply-imaged background galaxies at different redshifts have allowed the construction of very accurate total mass maps of the clusters to an unprecedented spatial resolution \citep[e.g.][]{Jauzac2014, Jauzac2015b, Balestra2016, Diego2016a, Annunziatella2017, Williams2017}. These data not only provide a unique possibility to study the high-redshift Universe, but also substructure in galaxy clusters \citep[e.g.][]{Jauzac2016b}.

Throughout this work, we adopt a standard cosmological model with the following parameters: $H_0$=70 km s$^{-1}$ Mpc$^{-1}$, $\Omega_m$=$0.3$ and $\Omega_\Lambda$=$0.7$. 

\section{Data}
\label{data}

In this section we describe the observational data we have used to explore distribution of the ICL and the X-ray emission of the galaxy clusters. In addition, we describe the gravitational models used to characterize the distribution of total mass in the clusters.


\subsection{HST near-IR data to characterize the ICL}

The primary data set used for this work is based in the HST images of the six HFF clusters (ID13495, PI: J. Lotz, \citealt[][]{Lotz2017})\footnote{\url{http://www.stsci.edu/hst/campaigns/frontier-fields/FF-Data}}. Details on the processing of the observational data from the six HFF clusters for the study of the ICL can be found in MT18 and are summarized below. As the ICL is more prominent at redder bands, in order to derive the bi-dimensional distribution of the ICL we used the HST \textsl{F160W} filter for each of the clusters. 

The data were directly retrieved from the HFF webpage\footnote{\url{http://www.stsci.edu/hst/campaigns/frontier-fields/FF-Data}}. For the ACS/WFC (Advanced Camera for Surveys Wide Field Camera, \citealt{Ford1998}) and WFC3 (Wide Field Camera 3, \citealt{MacKenty2008}) cameras, flat fields are claimed to be accurate to better than $1\%$ across the detector. The \textsl{F160W} images used here are drizzled science images with pixel size $0\farcs06$ (the native WFC3 pixel is closer to twice this value: $0\farcs13$).
The sky correction was carefully done by subtracting a constant measured in $\sim 30$ apertures of $r$ = $25$ pix ($1\farcs5$), well separated from any sources or diffuse light to minimize contamination. Kolmogorov-Smirnov tests were performed to confirm that the measured background followed a Gaussian distribution, with p-values $>0.05$.

Once the images are corrected for sky background, the next task is to identify the ICL on the images. This is a highly non trivial step as it requires to mask all the foreground and background objects on the images which are not part of the cluster and also all the cluster galaxies that are not the brightest cluster galaxy (BCG). Observations \citep[e.g.][]{Krick2007} have shown that the ICL is more centrally concentrated than the galaxies of the cluster implying that this light is formed via the mergers that build up the BCG. Consequently, there is no clear differentiation between the outskirts of the BCG and the extended ICL \citep[see][]{Mihos2016}. Therefore, we did not attempt to separate both components.
Even though in MT18 we performed extensive masking of sources that are not the BCG and ICL, for the current work we further increased the masks to get rid of any remaining low surface brightness light in the periphery of the galaxies that might contaminate the ICL.

A further complication in the analysis of the ICL distribution is that the HFF clusters are in the process or have experienced recent merging \citep[][]{Lotz2017}. Consequently, identifying the BCG is a not straightforward task and in three of the galaxy clusters (A2744, M0416 and A370) we decided to label the two most massive galaxies as BCGs. In these three cases, we effectively have two centers for the ICL distribution. Finally, in order to reduce the noise, especially at larger distances from the BCG(s), we smoothed the background-subtracted images using a gaussian of $\sigma=15$ pix ($\sim0.9\arcsec$).

\subsection{X-rays imaging}

In order to compare the shapes of the X-ray emission of the cluster to the mass distributions, we retrieved Advanced CCD Imaging Spectrometer (ACIS) images from the \textit{Chandra} Data Archive\footnote{\url{http://cda.harvard.edu/chaser/}}. The spatial resolution for Chandra ACIS imaging is $\sim1$ arcsec. The images were downloaded already processed. The clusters were observed with ACIS-S in 'VFAINT' mode, except for A370 which was observed by ACIS-I in 'FAINT' mode. They were processed with CXC software using CalDB. The versions of the CXC software used for processing the X-rays images are: A2744: 8.4.4, M0416: 10.3, M0717: 10.1.1, M1149: 10.2.1, AS1063: 10.5 and A370: 8.1.1. For CalDB the versions used are: A2744: 4.4.9, M0416: 4.6.4, M0717: 4.6.4, M1149: 4.6.4, AS1063: 4.7.2 and A370:4.1.4a.

A summary of the \textit{Chandra} data used in this work including observation ID numbers, principal investigator (PI) and effective exposure times can be found in Table \ref{table:xrays}.

\begin{table}
\begin{center}
\begin{tabular}{@{}l@{}|c@{}|c@{}|c@{}}
Cluster                                          &  Obs. ID    & PI                & Exp. Time (ks)    \\ \hline
Abell 2744 (A2744)                    &      8477       & Kempner   &   45.89 \\
MACS J0416.1-2403 (M0416)  &     17313       & Jones        &   71.13\\
MACS J0717.5+3745 (M0717)  &     16305       & Murray     &    94.34\\
MACS J1149.5+2223 (M1149)  &      16306      & Murray     &    79.71\\
Abell S1063 (AS1063)               &      18611      & Kraft         &    49.46  \\
Abell 370 (A370)                       &      515	       & Garmire    &    88.03 \\ 
\end{tabular}
\end{center}
\caption{Summary of the ACIS observations from the Chandra telescope used to derive the bi-dimensional X-ray distributions. }
\label{table:xrays}
\end{table}

\subsection{Mass models based on gravitational lensing}

The primary science goal of the HFF is to use the clusters as gravitational lenses to push the limits of current observations and study the faintest and most distant galaxies \citep[][]{Lotz2017}. However, in order to interpret many of the properties of those lensed galaxies, reliable models of the distribution of the mass for each cluster are required. As part of the HFF Initiative, different lensing models of the mass distribution for each of the six clusters have been provided to facilitate the analysis of the data. Having access to deep observations and hundreds of background sources at different redshifts, they provide very accurate (0.2-11\%, \citealt{Lotz2017}) maps of the mass distribution of the clusters.

Cluster mass estimates determined by lensing are valuable because the method requires no assumption about the dynamical state or formation history of the cluster and also provides us with an independently measured shape of the underlying DM halo. 

To conduct our goals, we used the $\kappa$ maps defined in units of the lensing critical density at the redshift of the lens ($\kappa = \Sigma / \Sigma_{crit}$) defined by
\begin{equation}
\Sigma_{crit} = \frac{c}{4\pi G}\frac{D_S}{D_L D_{LS}}
\end{equation}
using the angular-diameter distances from observer to source $D_S$, observer to lens $D_L$ and lens to source $D_{LS}$\footnote{\url{http://archive.stsci.edu/prepds/frontier/lensmodels/webtool/hlsp\_frontier\_model\_lensing\_primer.pdf}}. The models are scaled to $D_{LS}$/$D_S$ = 1. The lensing models are retrieved from MAST\footnote{\url{https://archive.stsci.edu/prepds/frontier/lensmodels/}} and summarized in Table \ref{table:mass}. 
We therefore made use of the $kappa$ maps from all the available models including CATS \citep[][]{Jullo2009,Richard2014,Jauzac2015a,Jauzac2015b,Jauzac2016a,Limousin2016,Lagattuta2017,Mahler2018}, Diego \citep[][]{Lam2014,Diego2015a,Diego2015b,Diego2016a,Diego2016b,Diego2018}, GLAFIC \citep[][]{Oguri2010,Ishigaki2015,Kawamata2016,Kawamata2018},  Keeton \citep[][]{Keeton2010}, Merten \citep[][]{Merten2009,Merten2011},  Sharon/Johnson \citep[][]{Johnson2014}, Williams/GRALE \citep[][]{Liesenborgs2006,Grillo2015,Sebesta2016}, Brada\u c \citep[][]{Bradac2009,Hoag2016}, Zitrin-NFW \citep[][]{Zitrin2013,Zitrin2015}, Zitrin-LTM \citep[][]{Zitrin2012,Zitrin2015}, and \citet[][]{Caminha2017}.

\begin{table*}
\begin{center}
\begin{tabular}{c|ccccc}
Cluster    &     \multicolumn{5}{c}{Models}        \\ \hline
\multirow{ 2}{*}{A2744}     &        CATS v4.1              &  Diego v4.1                 &        GLAFIC v4         &  Keeton v4                   &           Merten v1     \\
                &  Sharon \& Johnson v4  &          Willians v4        &     Brada\u c v2           & Zitrin-NFW v3             &        Zitrin-LTM v1    \\
\multirow{ 3}{*}{M0416}     &        CATS v4.1               &  Diego v4.1               &        GLAFIC v4         &  Keeton v4                    &           Merten v1     \\
                &  Sharon \& Johnson v4  &          Willians v4        &  Brada\u  c v3          &   Zitrin-NFW v3              &     Zitrin-LTM v1       \\
	        & Caminha v4                  &                                    &                                  &                                     &                               \\
\multirow{ 2}{*}{M0717}     &        CATS v4.1              &  Diego v4.1                &        GLAFIC v3          &  Keeton v4                   &      Merten v1          \\
               &  Sharon \& Johnson v4   &          Willians v4        &  Brada\u c v1            &     Zitrin-LTM v1            &                                \\
\multirow{ 2}{*}{M1149}     &         CATS v4.1             &  Diego v4.1                &        GLAFIC v3         &  Keeton v4                   &       Merten v1          \\
                &  Sharon \& Johnson v4  &          Willians v4        &  Brada\u c v1            &   Zitrin-LTM v1              &                                 \\
\multirow{ 2}{*}{AS1063}   &          CATS v4.1             &  Diego v4.1                &        GLAFIC v4         &  Keeton v4                   &        Merten v1         \\
               &  Sharon \& Johnson v4   &          Willians v4.1      &  Brada\u c v1           &   Zitrin-NFW v1             &     Zitrin-LTM v1        \\
\multirow{ 2}{*}{A370}       &               CATS v4          &  Diego v4.1                  &        GLAFIC v4        &  Keeton v4                   &       Merten v1           \\
              &  Sharon \& Johnson v4   &          Willians v4.1       &  Brada\u c v1           &   Zitrin-NFW v1             &     Zitrin-LTM v1        \\
\end{tabular}
\end{center}
\caption{Summary of the gravitational lensing models used. }
\label{table:mass}
\end{table*}

\section{2D Shape Matching}

The aim of this work is to study how well the distribution of the diffuse ICL in clusters traces the underlying distribution of the total mass of the cluster (baryonic + DM). In order to do that, we have taken advantage of the gravitational lensing ($\kappa$ maps) models as a way to infer the ``true" distribution of mass in the cluster. The question now is how to quantify the similarity between the bi-dimensional distribution of the ICL in comparison with the mass maps derived from gravitational lensing. In this work, we follow a two steps approach. First, we visually explore whether the mass distribution and the ICL have the same form by exploring the shape of the isophote and isomass contours of both distributions. In a second step, we quantify the similarities between both distributions through the Modified Hausdorff distance.

\subsection{Contours}
\label{contours}

To explore whether the ICL follows the 2-D distribution of DM in the HFF clusters, we derived the isocontours for each of the different components we are studying: light (ICL) and mass. In addition, to compare with the hot gas component, we also extract the isocontours for the X-ray emission. In order to have a sensible comparison, the isocontours of each map were obtained at the same physical radial distances: $50$, $75$, $100$, $125$ and $140$ kpc. To measure a radial distance we need to define a centre. Because the location of the centre affects the location of the different radial distances, and therefore the shape of the contours, we expand on the choice of centres for each component.

In the case of the ICL, the obvious centre is the one provided by the position of the BCG (or BCGs). For the gravitational lensing mass models, we also use the peak(s) of the distribution of total mass (which roughly corresponds to the position of the BCG(s) in each cluster). In the case of the X-ray observations, the centre is the location of the peak of emission. This is done for a reason: the gas is a dissipative component and its distribution is clearly disturbed in the majority of the HFF clusters. Therefore, the choice of the peak of X-rays as the centre of the hot gas distribution has the aim of making a much fairer comparison in the disturbed cases. Nevertheless, we also tested our procedure by fixing the centre of the different distributions to the BCG(s) and the qualitative results of this paper do not change with the choice of the centres.

\begin{figure*}
  \centering
  \includegraphics[width=\textwidth]{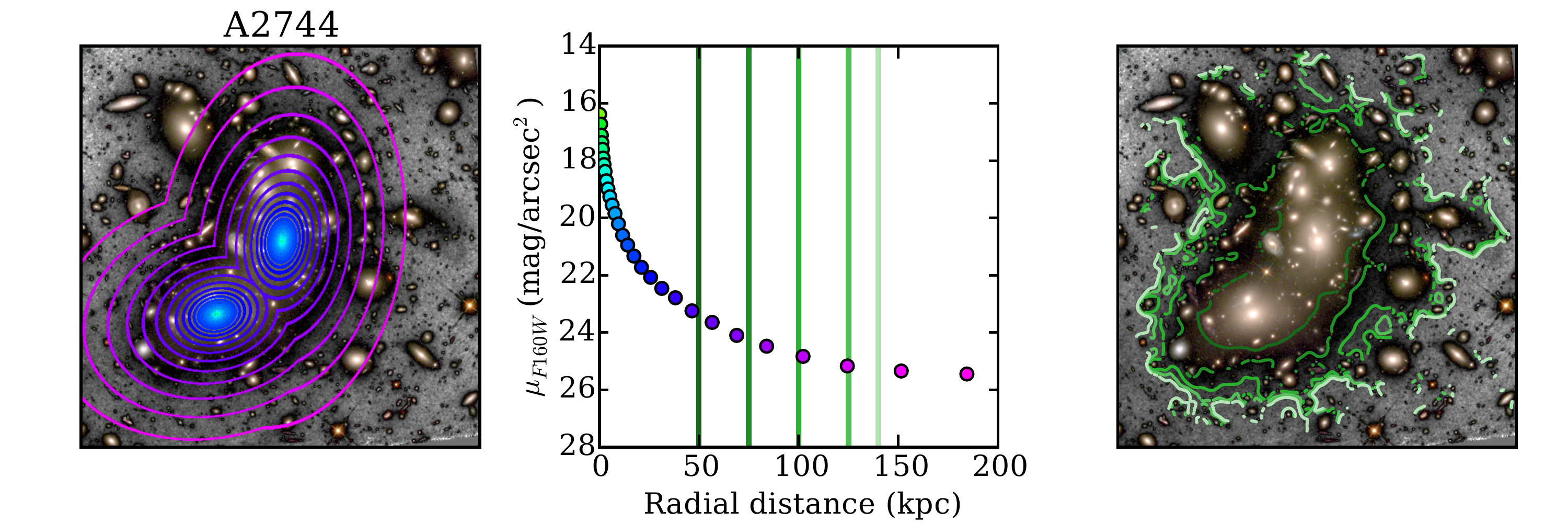}
  \caption{A description of the procedure for obtaining the isocontours of ICL, mass and X-rays. {\it Left panel:} RGB image of one of the HFF clusters with the different spatial regions in which the surface brightness profile is measured. The radial distance to each pixel in each radial bin is computed as the elliptical distance to its nearest BCG. {\it Middle panel:} \textsl{F160W} surface brightness profile of the cluster averaging the values in each radial bin. The green vertical lines correspond to the radial distances of $50$, $75$, $100$, $125$ and $140$ kpc where we infer the values to draw the contours. {\it Right panel:} Contours of equal surface brightness for the values obtained at the radial distances indicated in the middle panel.}
  \label{fig:plot_method}
\end{figure*}

Once the centres are chosen, we obtain the radial light/mass profiles of the ICL, X-ray emission and mass maps. To that end, we average the intensities in 16 logarithmic spaced bins from $0$ to $200$ kpc from the BCG(s). The rafial distance to each pixel on the images is computed as the elliptical distance to its nearest centre (BCG) as seen in the left panel of Fig. \ref{fig:plot_method}. The morphological parameters (i.e., axis ratio and position angle) for these galaxies are given by \texttt{SExtractor} (see MT18 for further details). This procedure is illustrated in Fig. \ref{fig:plot_method} for the ICL. Using this radial profile, we interpolated the intensities of the ICL/mass/X-ray profiles at the radial distances $50$, $75$, $100$, $125$ and $140$ kpc.
Once the intensities at different radial distances are characterised, we build the contours on the different maps that correspond to such intensities using the function \texttt{contour} in \texttt{matplotlib}. The \texttt{contour} function provides and draws isocontour lines at different given values of the image, in this case the intensities for the 5 radial distances.
The contours for the 5 distances for each cluster and each component are shown in Fig. \ref{fig:contourall}. Left panels show the contours for the ICL (green), middle panels for the total mass of the cluster (blue) and right panels for the X-ray emission (red). To make a proper comparison of the contours, we have applied the same masking that we derive in the \textsl{F160W} filter to the rest of the maps. 

The mass models shown in the middle panels of Fig. \ref{fig:contourall} are the models representative of the average shape of all the models at all radius for each cluster (i.e. the mass models used are: A2744: GLAFIC v4, M0416: Sharon \& Johnson v4, M0717: CATS v4.1, M1149: CATS v4.1, AS1063: Diego v4.1, A370: Keeton v4).

\begin{figure*}
\centering
  \subfigure{\includegraphics[width=\textwidth]{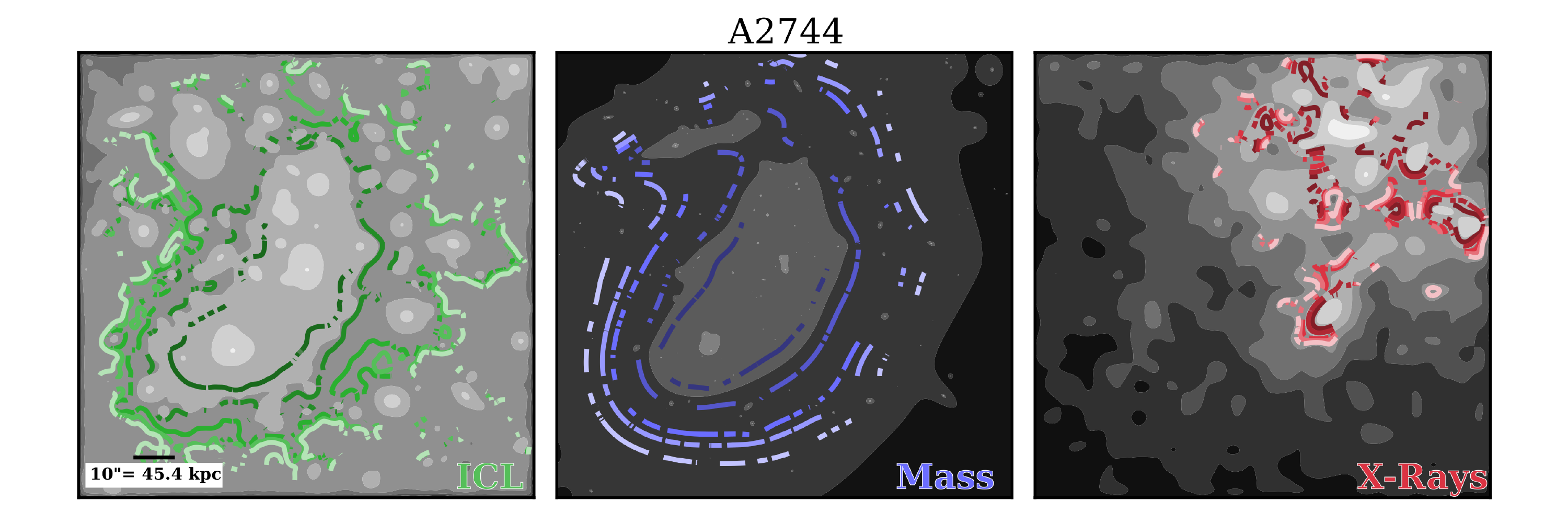}}
  \subfigure{\includegraphics[width=\textwidth]{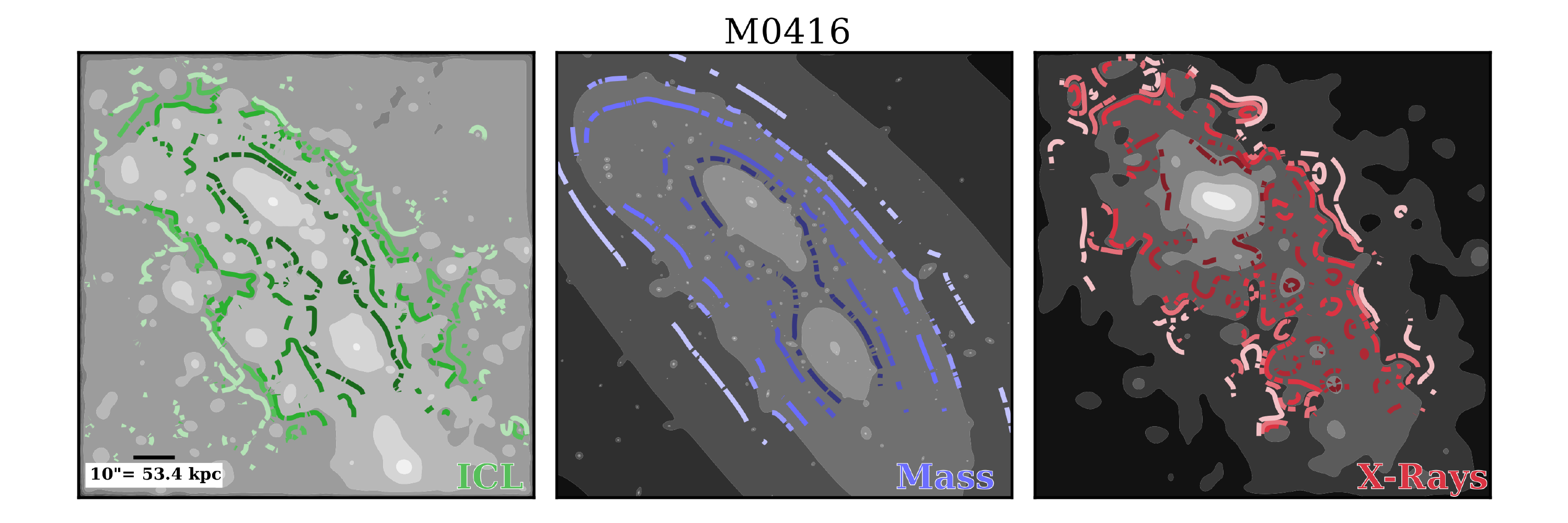}}
  \subfigure{\includegraphics[width=\textwidth]{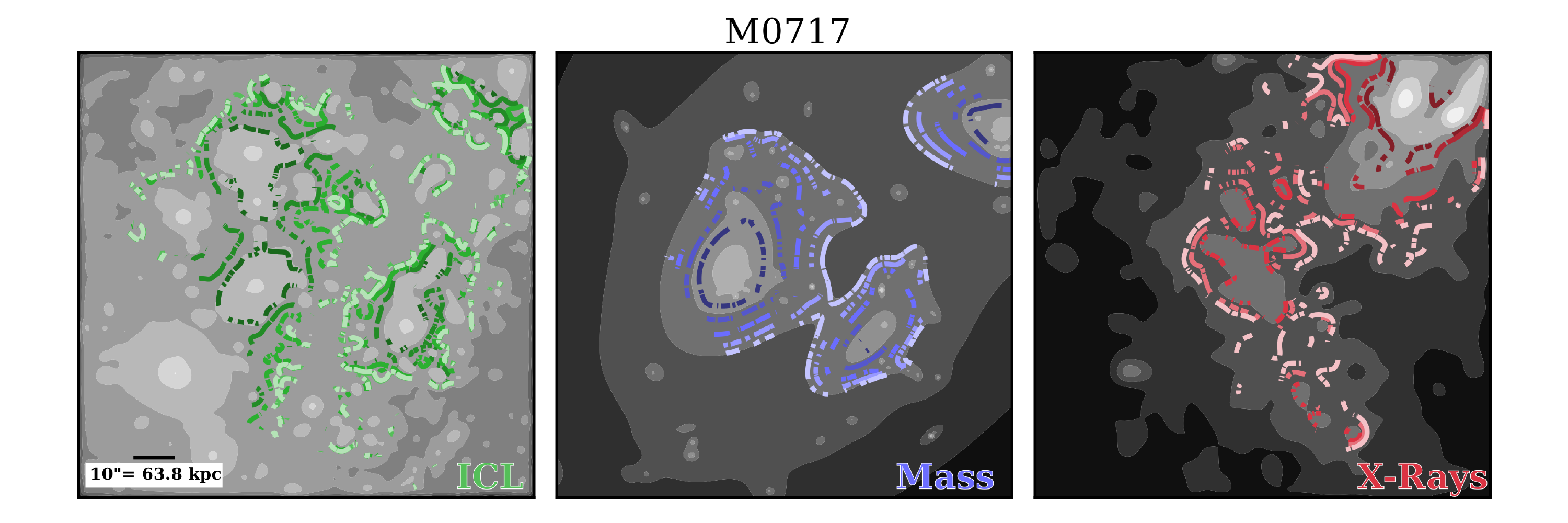}}
  \caption{Comparison of the distribution of the ICL from the F160W images (left panels, green), the mass distribution of the cluster based on its gravitational lensing (middle panels, blue) and the distribution of the hot gas X-ray emission (right panels, red) for each of the FF clusters. The size of the FOV is $110\times 110$ arcsec$^2$ in each cluster. We overplotted the contours for 5 different distances: 50 (darkest colors), 75, 100, 125 and 140 (lightest colors) kpc.} \label{fig:contourall}
\end{figure*}

\setcounter{figure}{1}    

\begin{figure*}
  \centering
  \subfigure{\includegraphics[width=\textwidth]{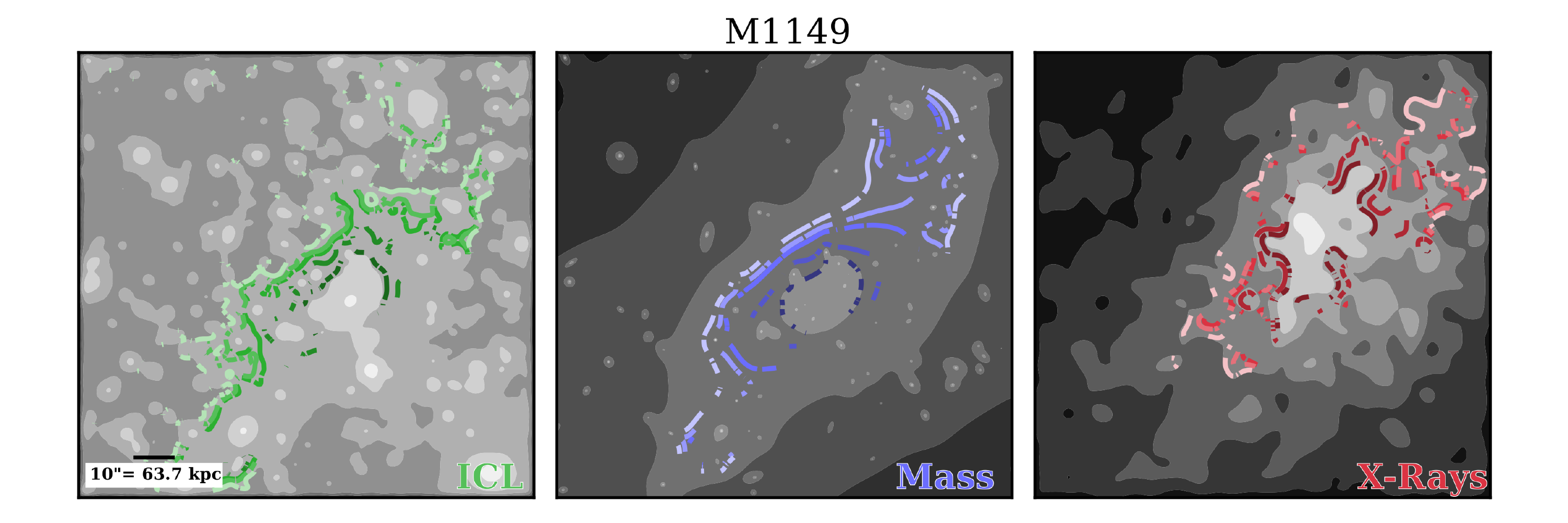}}
  \subfigure{\includegraphics[width=\textwidth]{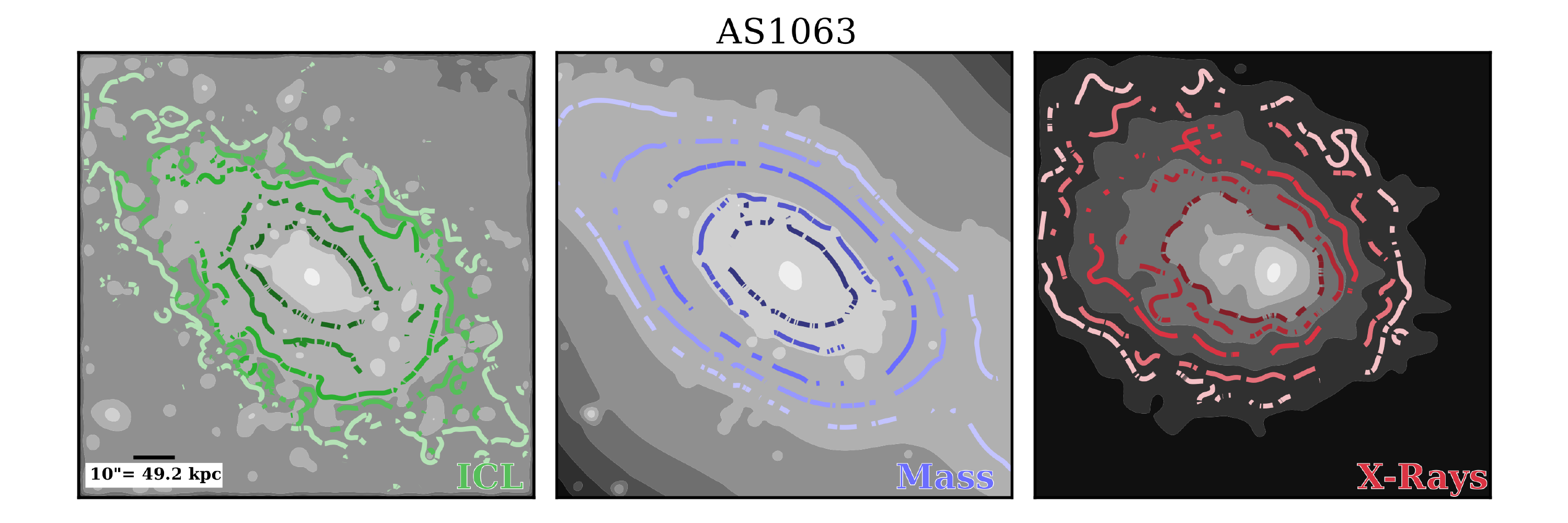}}
  \subfigure{\includegraphics[width=\textwidth]{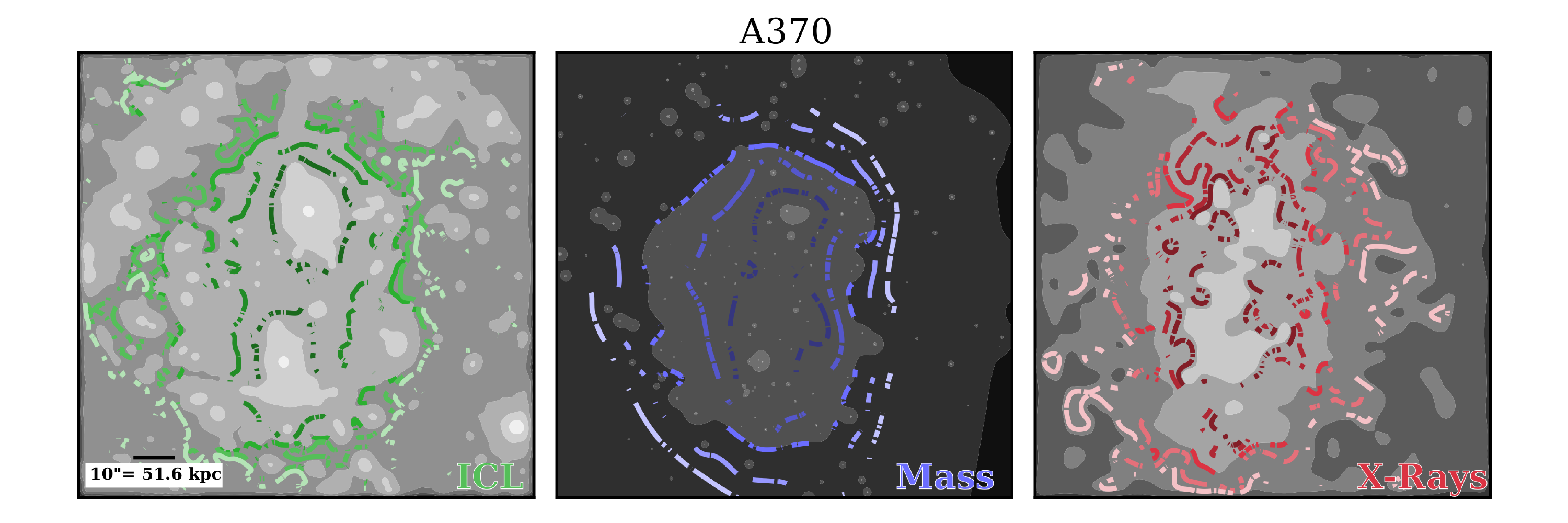}}    
  \caption{Continued...}
\end{figure*}

\begin{figure*}
\vspace{-10pt}
  \centering
  \begin{tabular}{|@{\hspace{2pt}}c@{\hspace{-25pt}}c@{}}
  \subfigure{\includegraphics[width=0.56\textwidth]{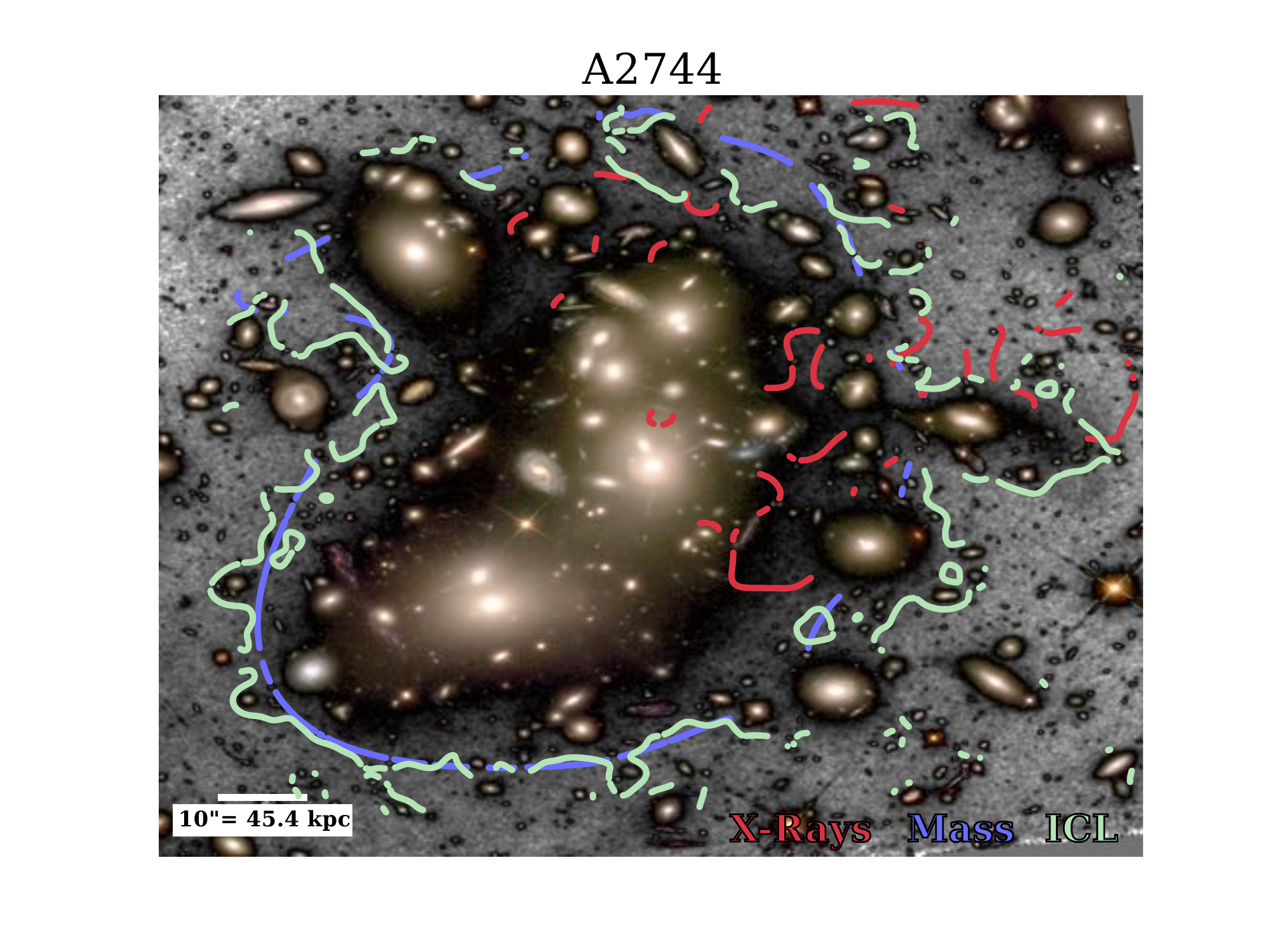}} &
  \subfigure{\includegraphics[width=0.55\textwidth]{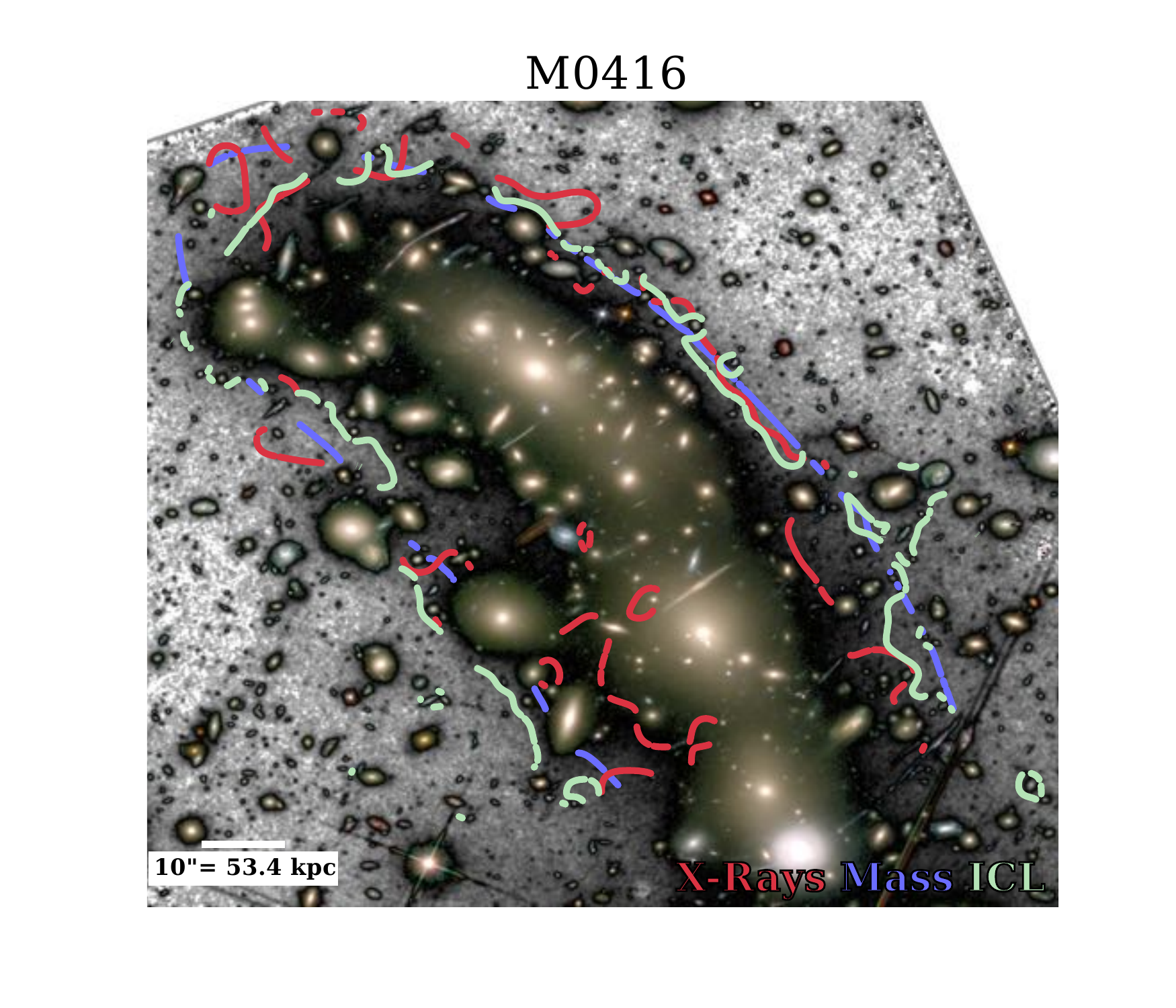}} \\[-33pt]
  \subfigure{\includegraphics[width=0.55\textwidth]{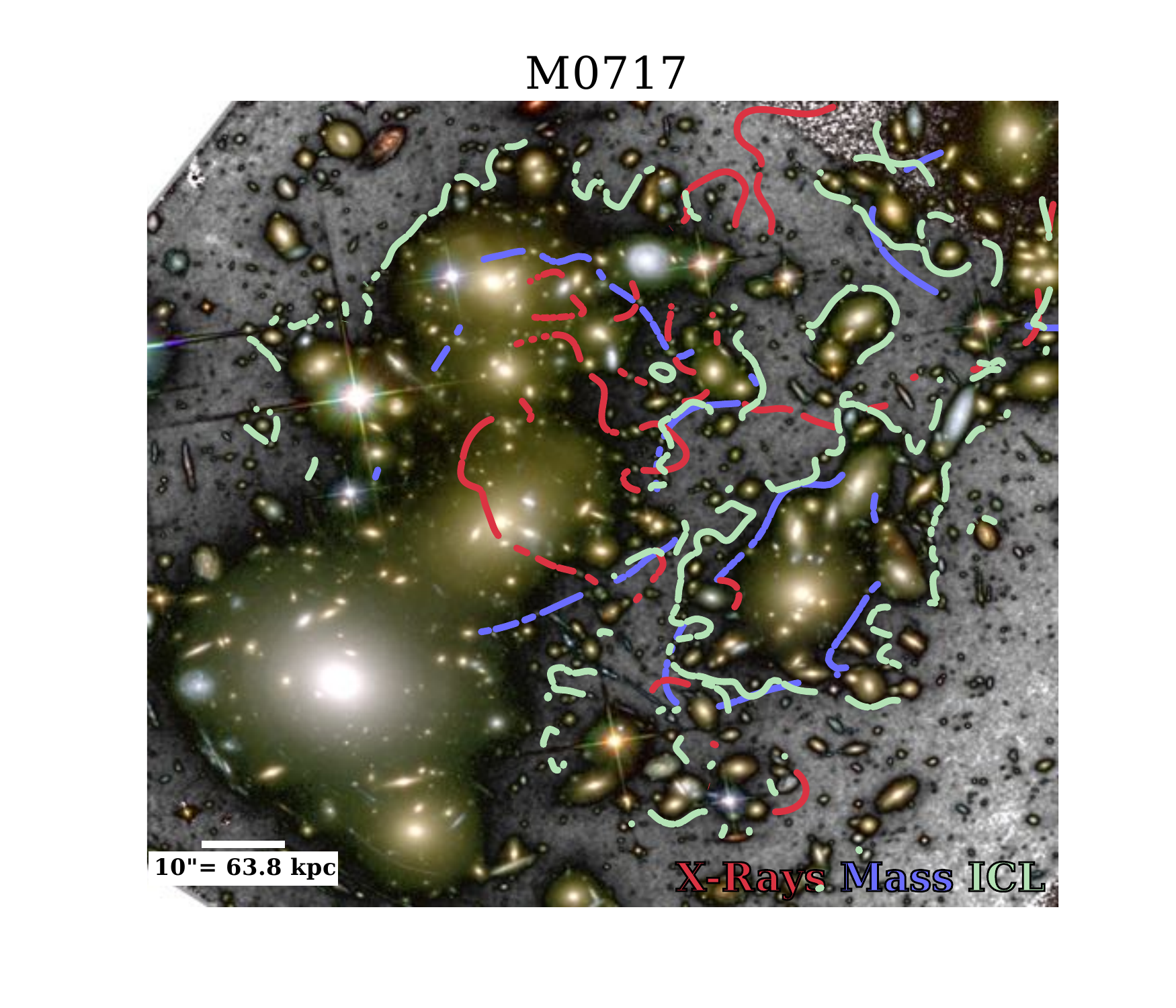}} &
  \subfigure{\includegraphics[width=0.55\textwidth]{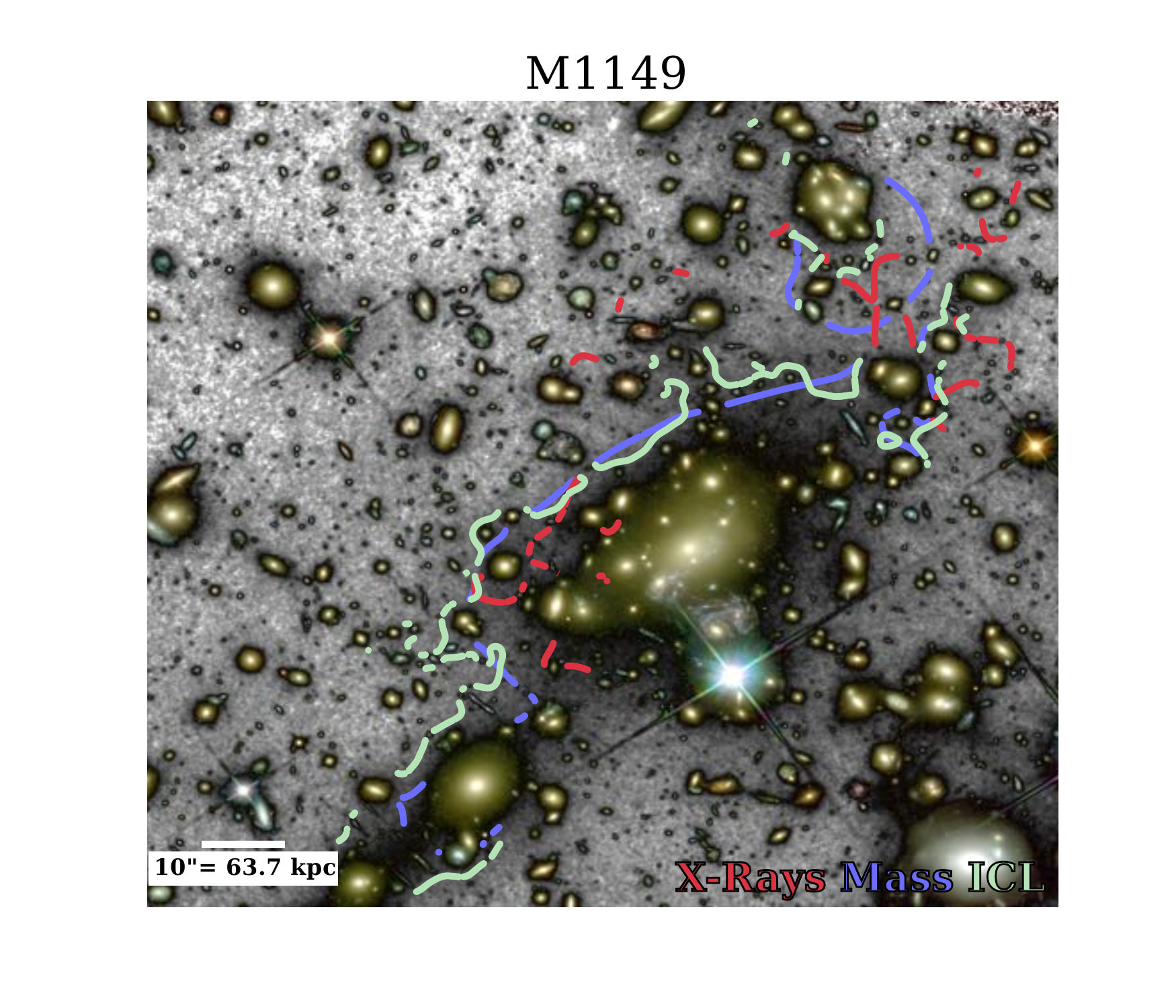}} \\[-33pt]
  \subfigure{\includegraphics[width=0.55\textwidth]{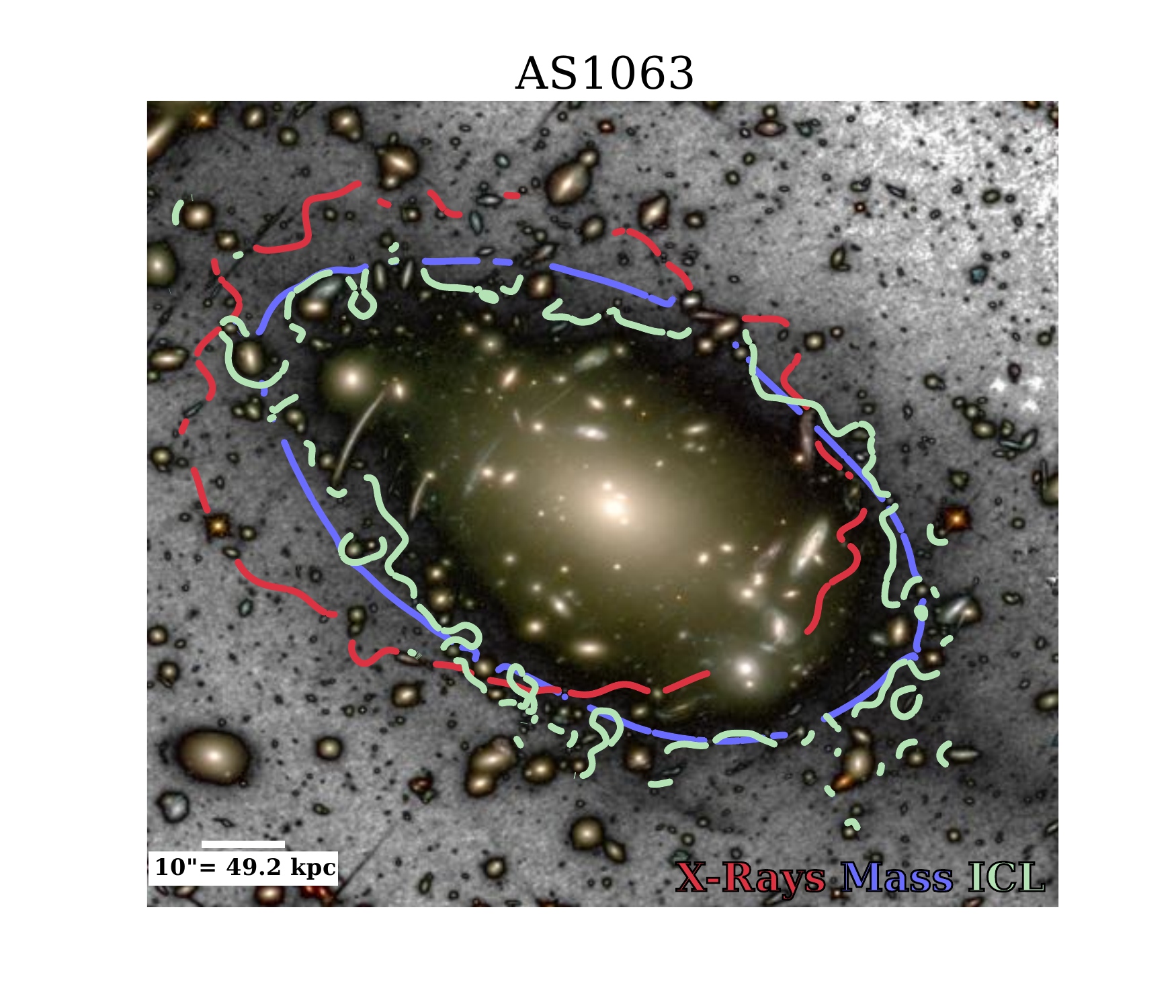}} &
  \subfigure{\includegraphics[width=0.55\textwidth]{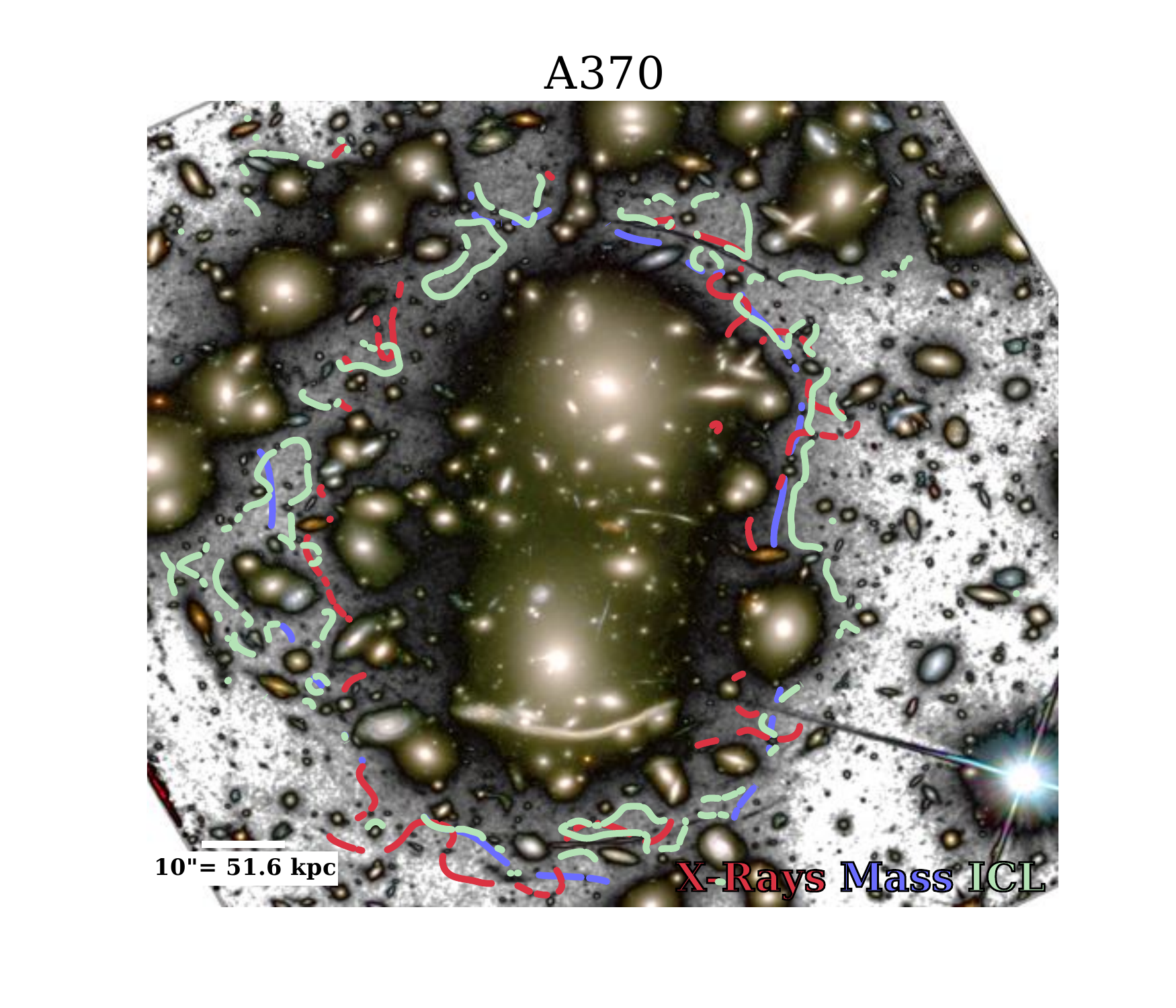}} \\
  \end{tabular}
\caption{RGB images of the HFF clusters with the contours of ICL (green), X-rays (red) and Mass (blue) at a radial distance of $125$ kpc overplotted. The RGB images are a combination of the F606W, F814W and F125W bands whereas a black and white F160W image is used for the background.} \label{fig:contourrgb}
\end{figure*}

For ease of comparison, we plotted in Fig. \ref{fig:contourrgb} the contours of the ICL (green), mass (blue) and X-rays (red) corresponding to a radial distance of $125$ kpc. The underlying image is a composite of an RGB color image created using the F606W, F814W and F125W filters and a black and white F160W image for the background. 

A first visual inspection shows that the bi-dimensional distribution of the ICL is comparable to the mass distribution of the clusters. However, this is not always the case for the X-rays. Only in one of the six cases, A370, the X-ray distribution is in agreement with the total mass distribution and, also, the ICL. This observed difference between the X-ray emission and the distribution of total mass is expected. The HFF clusters are undergoing or have undergone recent mergers. In these cases, the hot gas that produces the X-ray emission experiences ram pressure and is slowed creating an offset between the DM and the X-ray emission. As mentioned in the Introduction, a well-known example of this is the Bullet Cluster where the X-ray and DM distribution are different \citep[][]{Clowe2004, Markevitch2004}.

\subsection{Modified Hausdorff Distance}

Admittedly, a visual inspection is not an ideal quantifier of the resemblance amongst the three sets of data. In order to measure the similarities between the contours of the different maps of the clusters, we chose a metric used in shape matching: the Modified Hausdorff distance (MHD). The MHD is a measure of how far two subsets are from each other. 
Given two sets of a metric space, the Hausdorff distance (HD), named after Felix Hausdorff, is defined as the maximum distance of all the distances from a point in one set to the closest point in the other set. For two sets on points $X = \{x_1, x_2, \cdots x_{N_x}\}$ and $Y = \{y_1, y_2, \cdots y_{N_y}\}$, the HD is

\begin{equation}\label{eq:eq1}
HD(X,Y) = max(d(X,Y), d(Y,X))
\end{equation}

\begin{equation}\label{eq:eq2}
d(X,Y) = \max\limits_{x_i \in X} \min\limits_{y_j \in Y}  ||x_i-y_j||
\end{equation}

The smaller the value of HD, the more similar the two point sets are. Therefore, this distance has applications in the context of shape matching \citep[][]{Huttenlocher1993}. However, as the Hausdorff distance is set by the maximum distance among two sets of points it can be very sensitive to outliers even when the objects are fairly similar \citep[e.g.][]{Dubuisson1994}. For that reason, \citet[][]{Dubuisson1994} introduced the MHD \citep[see also][]{Huttenlocher1993}. The MHD changes Eq. \ref{eq:eq2} for:

\begin{equation}
d(X,Y) = \frac{1}{N_X}\sum\limits_{x_i \in X} \min\limits_{y_j \in Y} ||x_i-y_j||
\end{equation}
where N$_X$ is the number of elements in the set of points X. The MHD is more robust to outliers than the original HD form. Another interesting property of the MHD is that its value increases monotonically as the difference between the two sets of points increases.

We can interpret the MHD as the mean distance difference between two sets of data. In our case we are interested in exploring how well a given tracer (X-ray or ICL) follows the distribution of the total mass of the cluster. Consequently, the MHD will give us the mean difference in distance (in kpc) between the tracer and the mass model. For instance, a MHD value around $20$ kpc would mean that the average departure of the ICL contours from the mass contours is of the order of the size of the Milky Way \citep[][]{Goodwin1998}.

\section{Results}

\begin{figure*}
  \centering
  \subfigure{\includegraphics[width=0.45\textwidth]{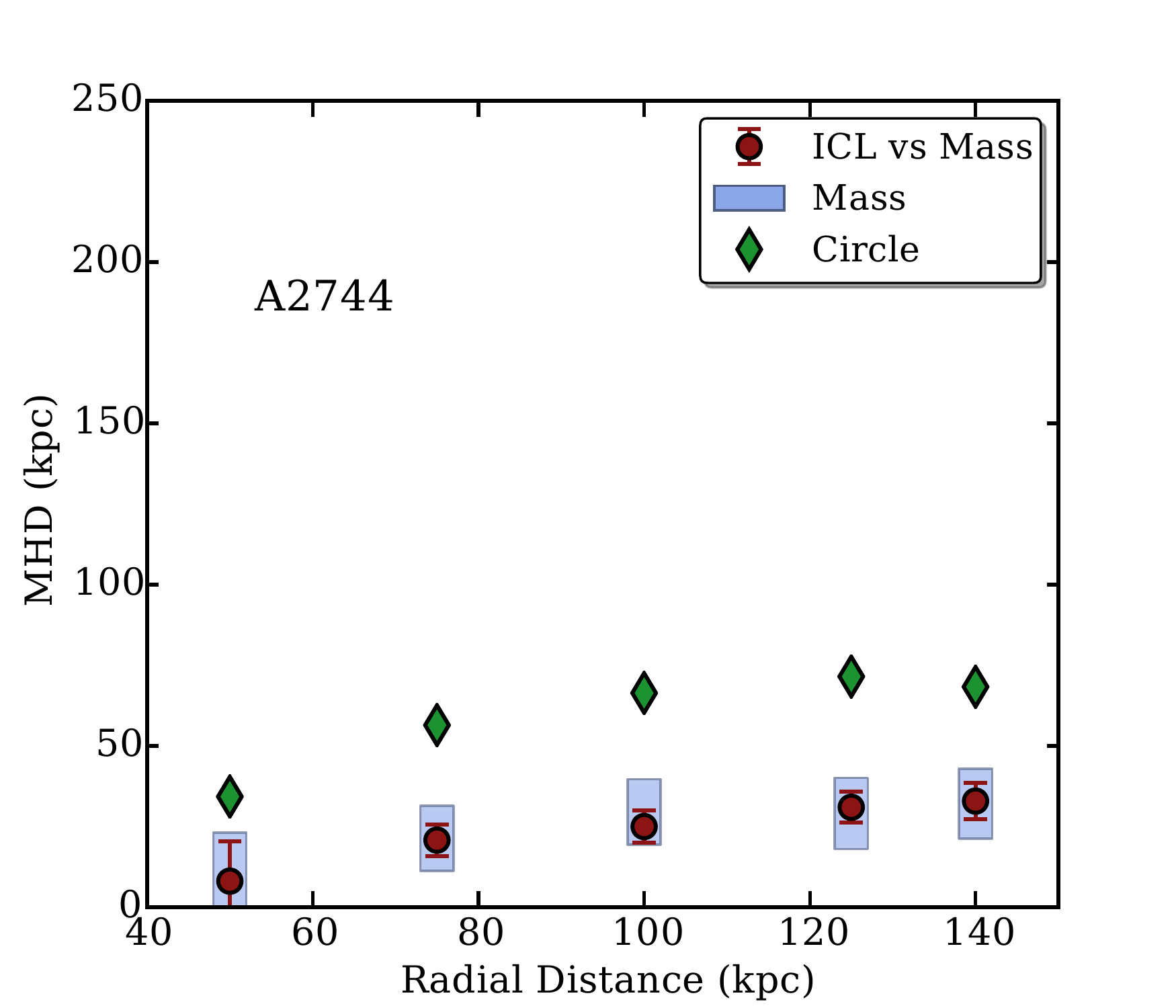}}
  \subfigure{\includegraphics[width=0.45\textwidth]{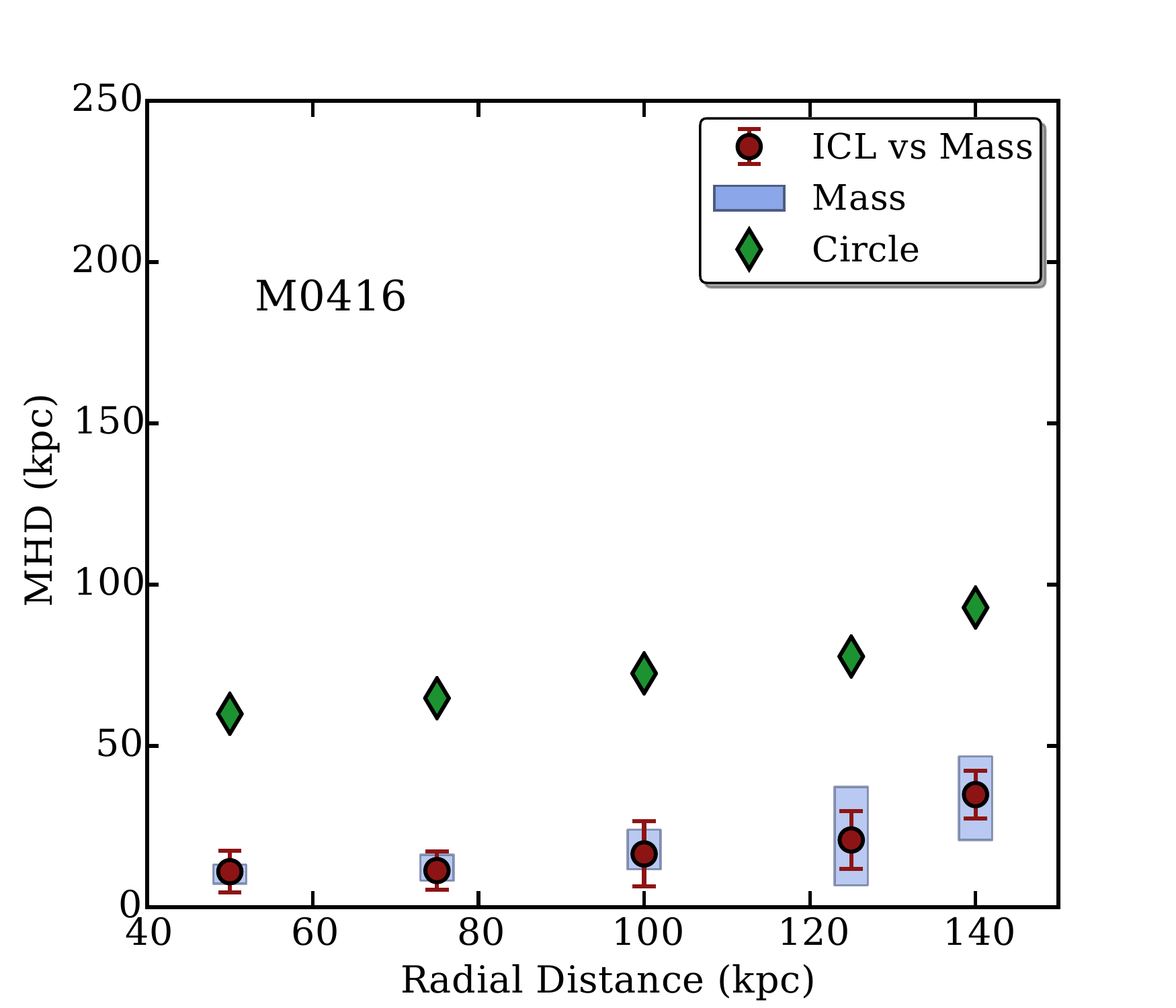}}
  \subfigure{\includegraphics[width=0.45\textwidth]{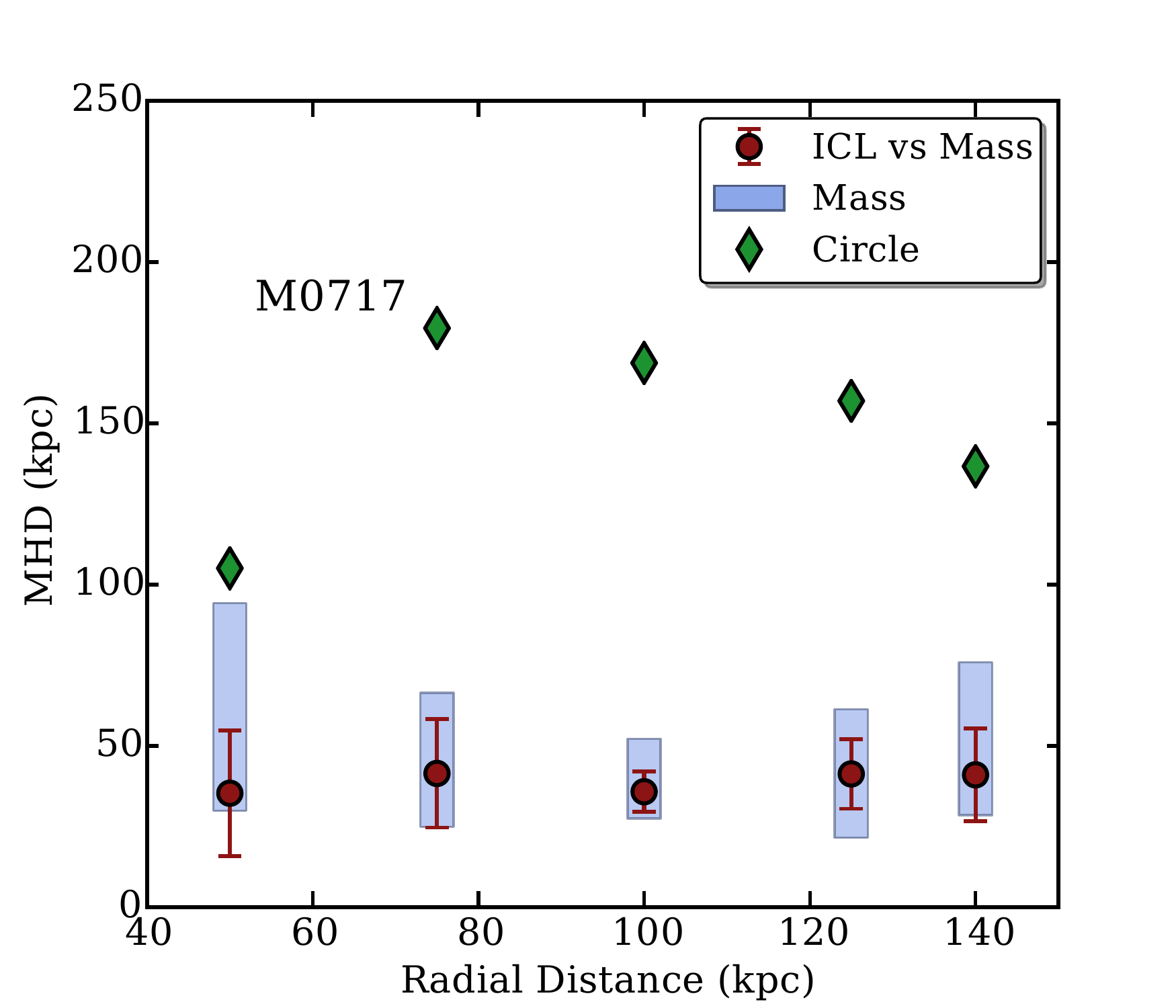}}
  \subfigure{\includegraphics[width=0.45\textwidth]{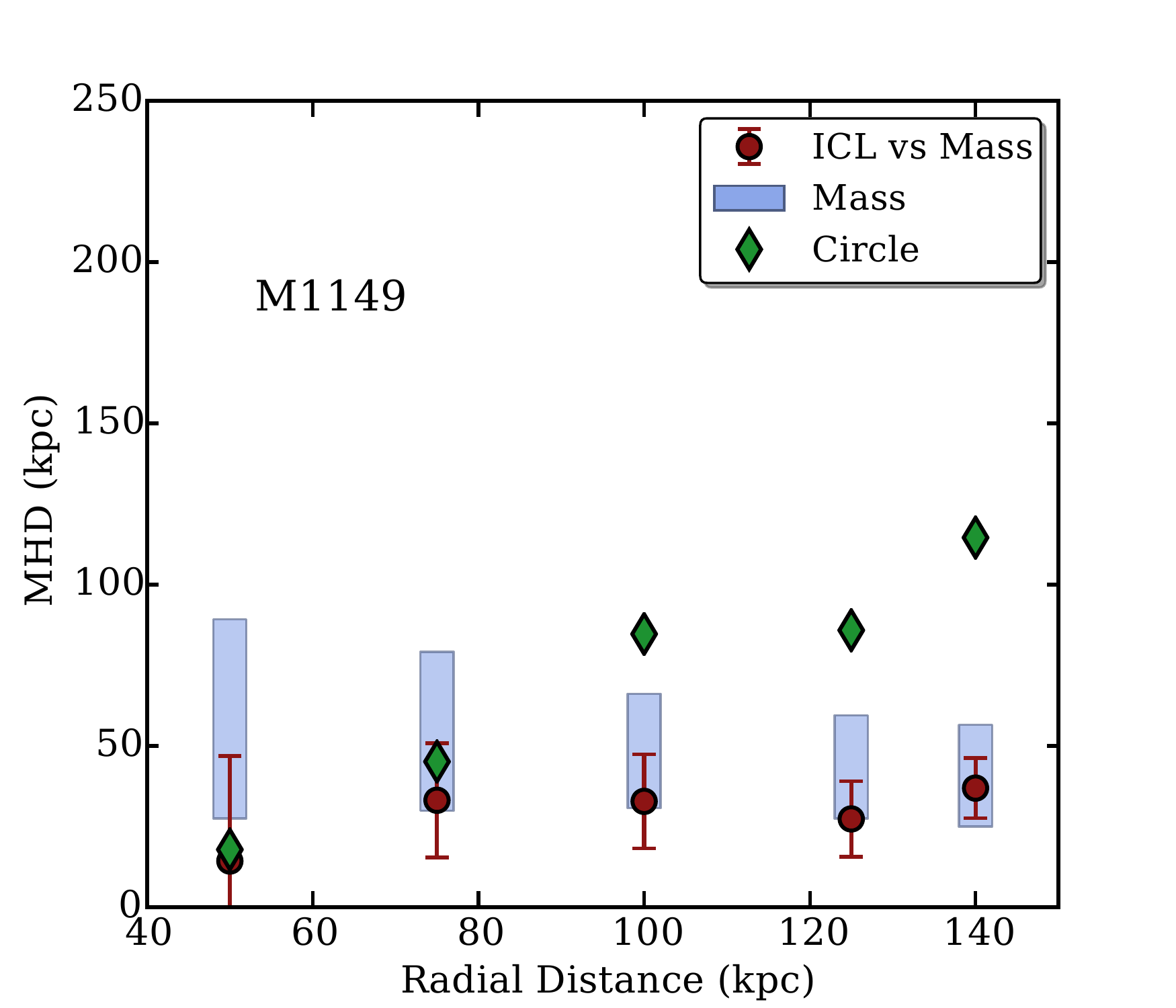}}
  \subfigure{\includegraphics[width=0.45\textwidth]{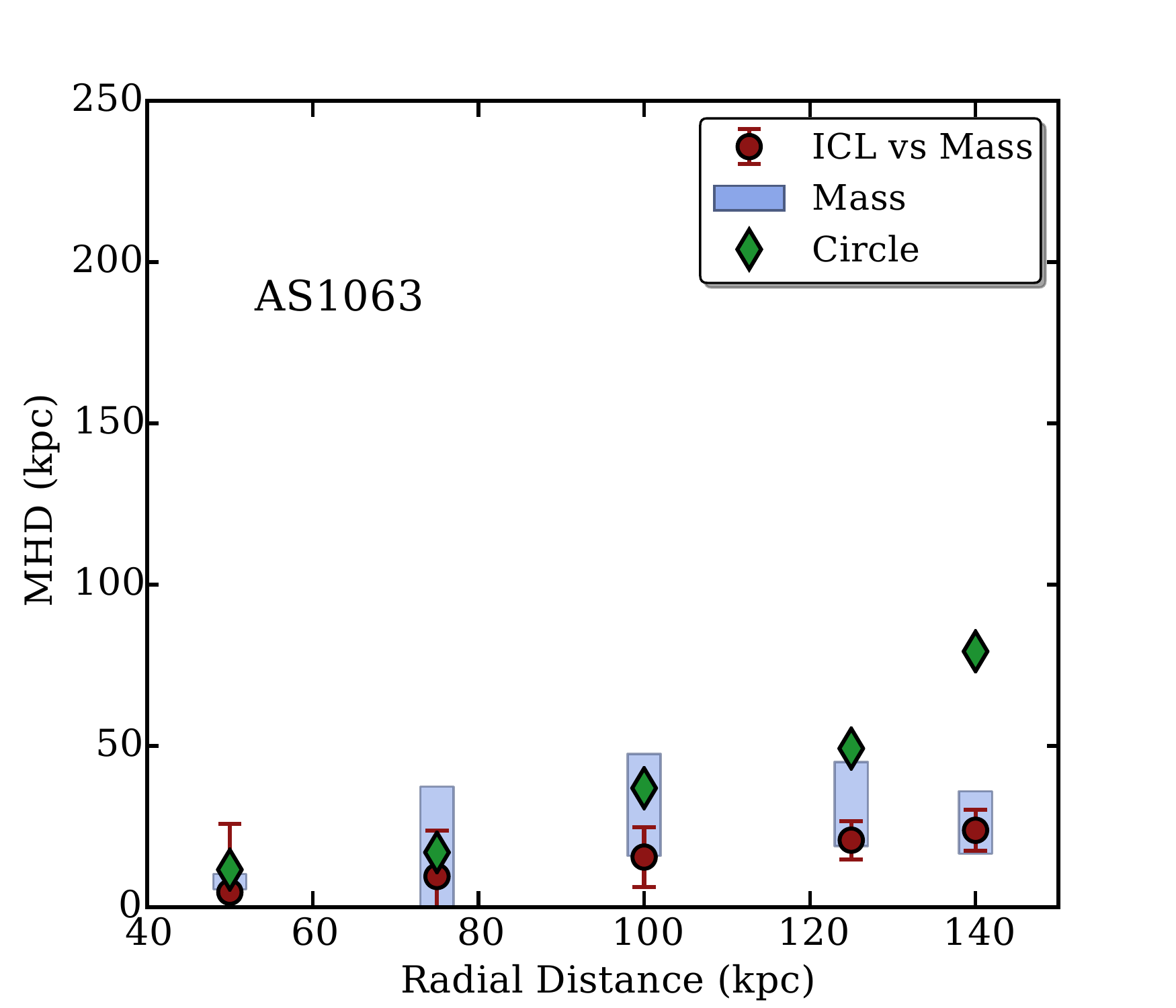}}
  \subfigure{\includegraphics[width=0.45\textwidth]{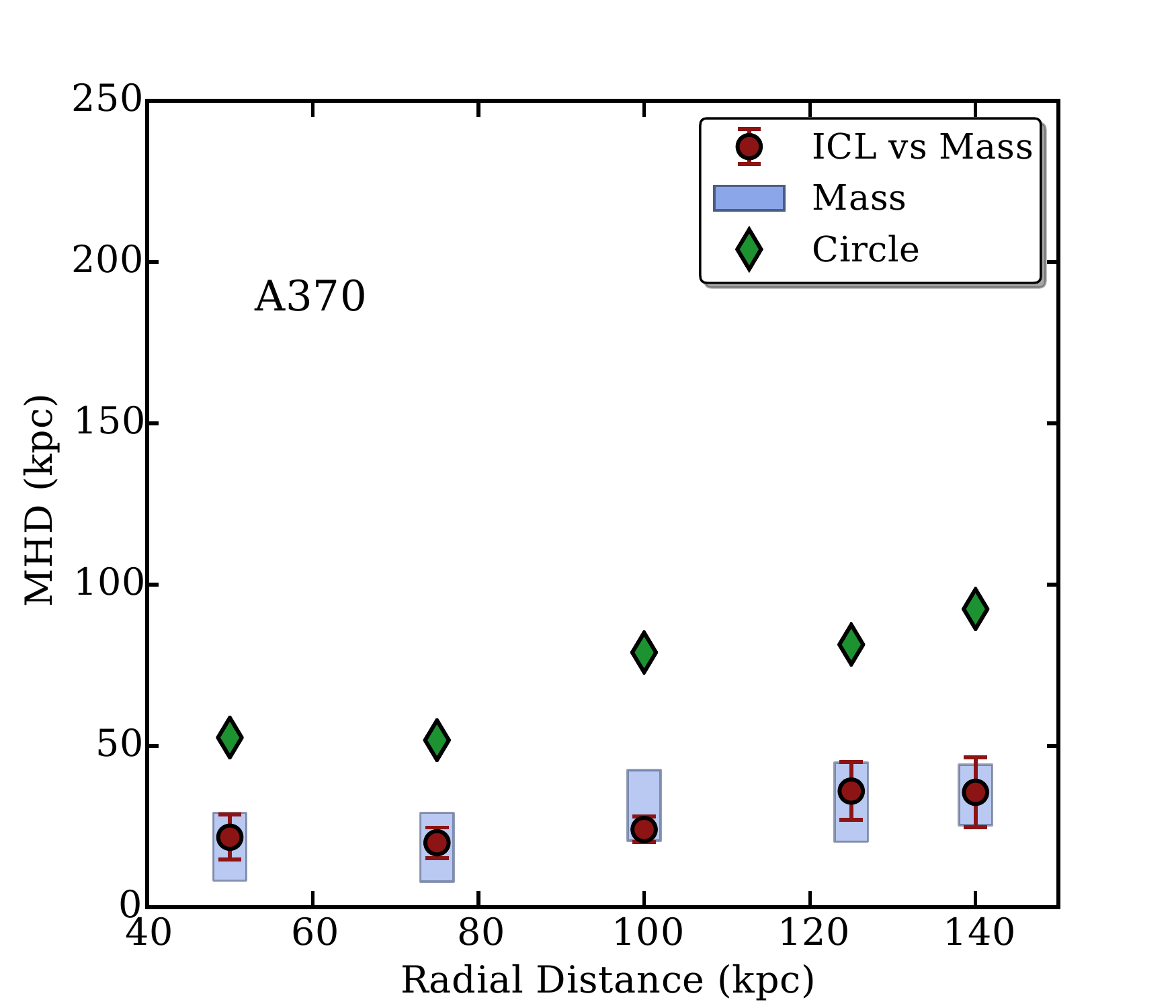}}
 
  \caption{Modified Hausdorff Distance (MHD) between the ICL of the HFF clusters and the different mass models from gravitational lensing. The blue rectangles enclose the $68$\% of the measured MHDs among the mass models themselves. The red filled circles represent the median values of the MHD of the ICL when compared with all the mass models. As a visual guide, we have also plotted the value of the MHD when comparing the ICL contour with a circle at each given radius (green diamonds). The departure of the diamonds from the red circles indicate an increasing ellipticity of the ICL distribution.}\label{fig:mhd}
\end{figure*}

In Fig. \ref{fig:mhd}, we plotted the MHD between the ICL and the mass maps as a function of the radial distance for each of the HFF clusters. The red circles represent the median\footnote{We use the resistant median i.e. trimming away the $5\sigma$ outliers.} value of the MHD at each distance resulting from the comparison between the contours of the ICL and the different mass maps for each of the models listed in Table \ref{table:mass}. The error bars represent the 1$\sigma$ scatter. The median values and errors are given in Table \ref{table:mhd_icl}. The process of converting the lensing constraints into matter distribution is not unique, and therefore, different models use different methodologies and also other tracers of the gravitational potential of the cluster to constrain the degeneracies (X-rays, galaxy kinematics, etc., see \citealt{Meneghetti2017} for a review of the different methodologies). \citet{Priewe2017} showed that the magnification maps based on different lens inversion techniques differ from each other by more than their statistical errors due to degeneracies. This difference can be understood as the intrinsic uncertainty in the mass distributions of the clusters. Accordingly, in order to evaluate how reasonable is the similarity among the ICL and the mass models, it is necessary to quantify what is the typical MHD between the different mass models themselves for a given cluster. The blue rectangles in Fig. \ref{fig:mhd} represent the $1\sigma$ uncertainty of the MHD among the mass models (enclosing $68\%$ of them\footnote{Note that the middle panel of Fig. \ref{fig:contourall} shows the gravitational lensing model with the MHD closer to the median MHD computed among models at all radii.}). The median values and uncertainties among the different mass models are given in Table \ref{table:mhd_models}.

In those cases where the MHD between the ICL and the mass models is similar to the typical MHD among the different mass models implies that the ICL is basically identical in shape to the isomass contours provided by the different mass models. In other words, the ICL distribution is fully compatible (within the uncertainties) with the mass distributions of the lensing models. In all the cases there is an excellent agreement between the ICL distribution and the distribution of total mass. The average MHD between the ICL and the mass models and the mass models themselves is around 25 kpc for most of the clusters. This means that the average difference between the contours is around the size of a Milky Way-like galaxy. It is noteworthy that the two higher redshift clusters, M0717 and M1149 exhibit larger uncertainties and higher MHD ($\sim$50 kpc) at all radii. This is caused by a lack of spectroscopically confirmed lensed systems and images, and therefore less constrained mass models \citep[e.g.][]{Limousin2016,Jauzac2016a,Natarajan2017,Williams2017}. In the following, we will discuss in more detail each cluster.

\begin{figure*}
  \centering
  \subfigure{\includegraphics[width=0.45\textwidth]{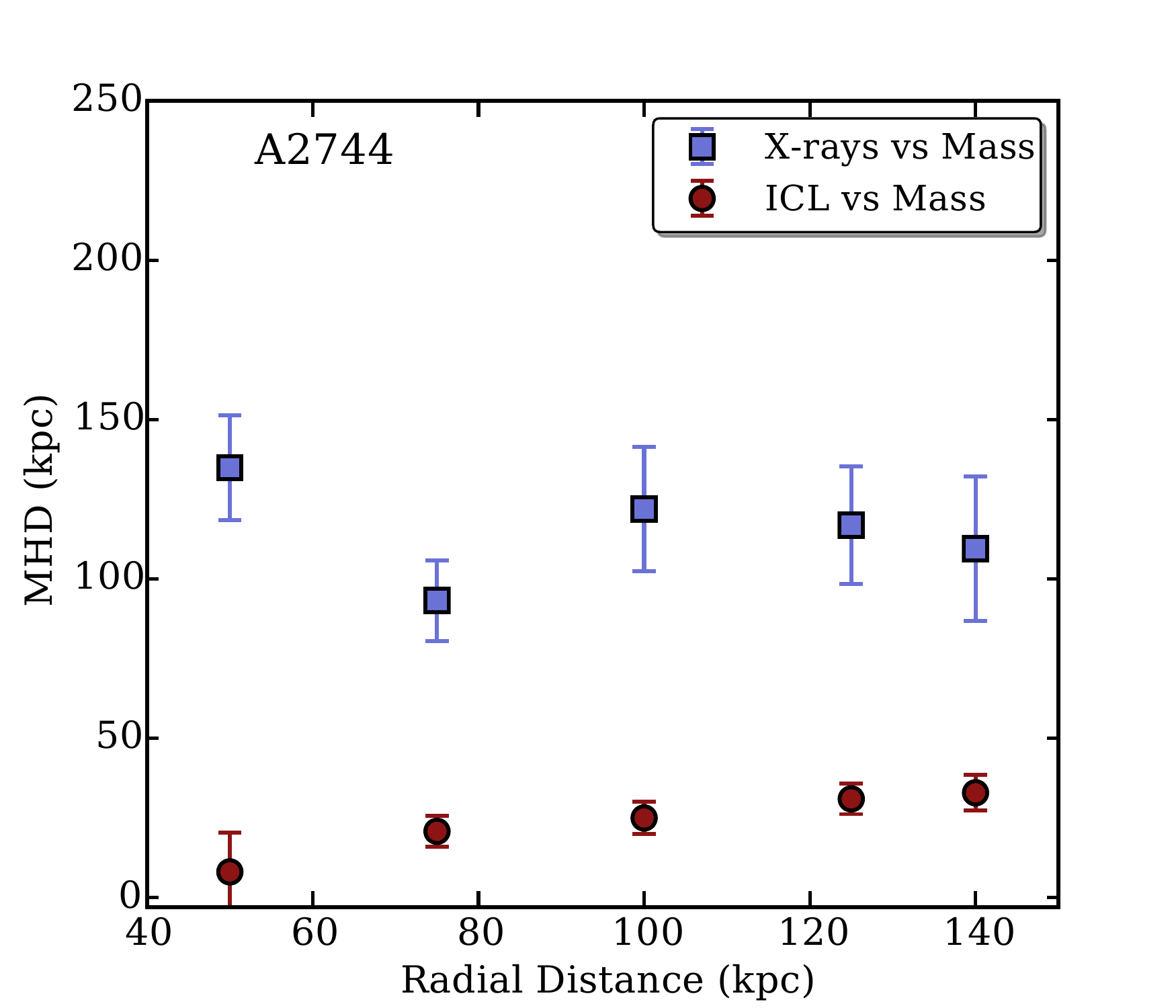}}
  \subfigure{\includegraphics[width=0.45\textwidth]{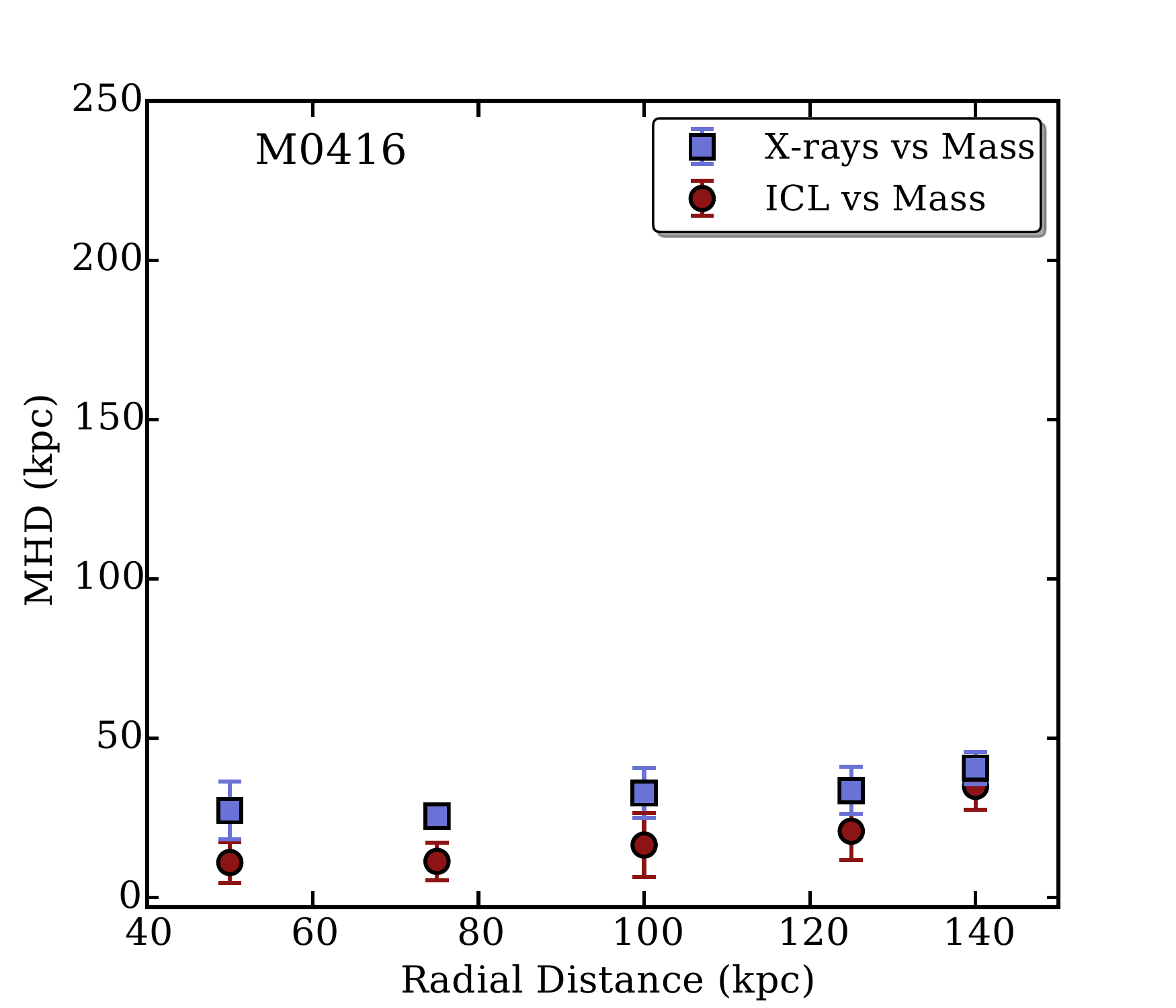}}
  \subfigure{\includegraphics[width=0.45\textwidth]{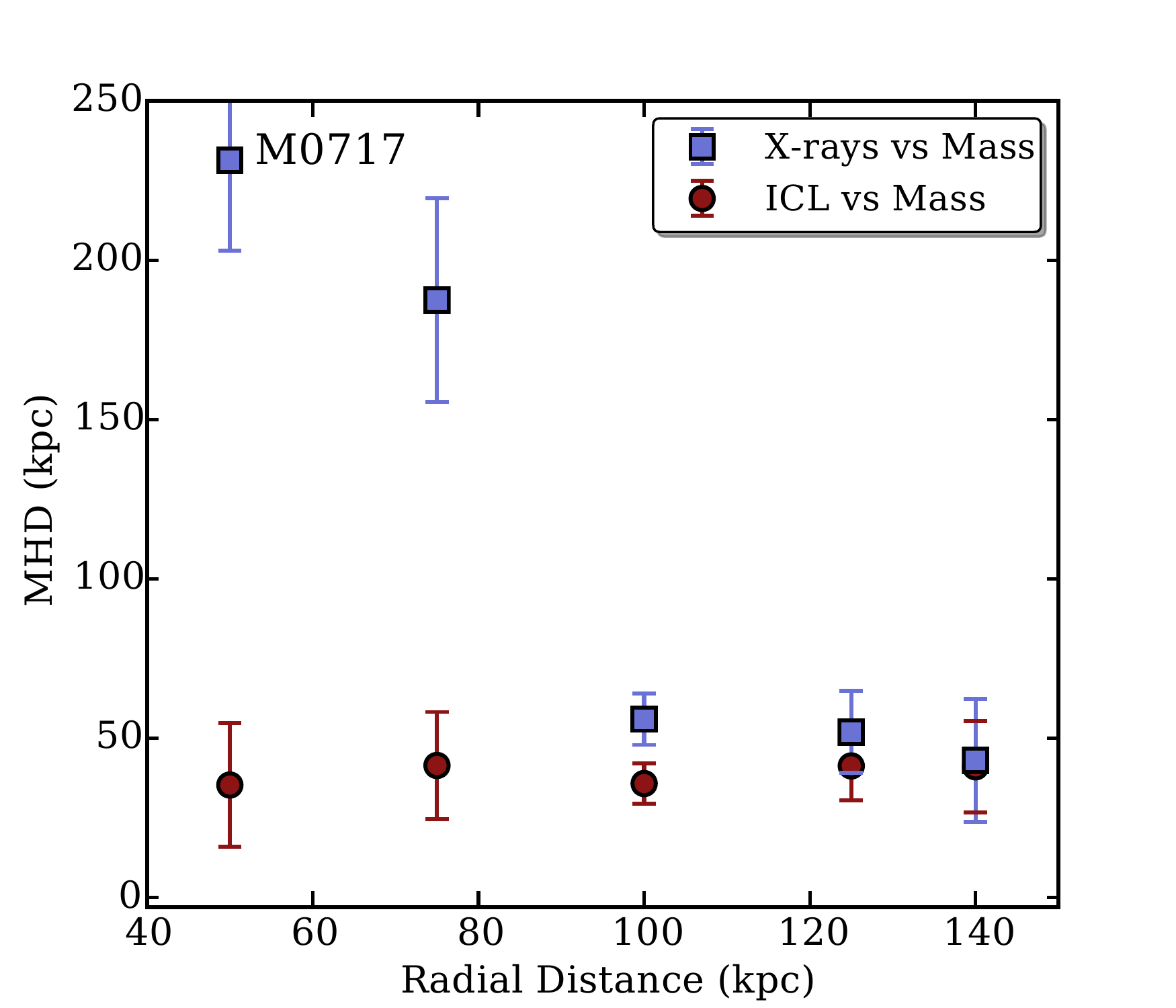}}
  \subfigure{\includegraphics[width=0.45\textwidth]{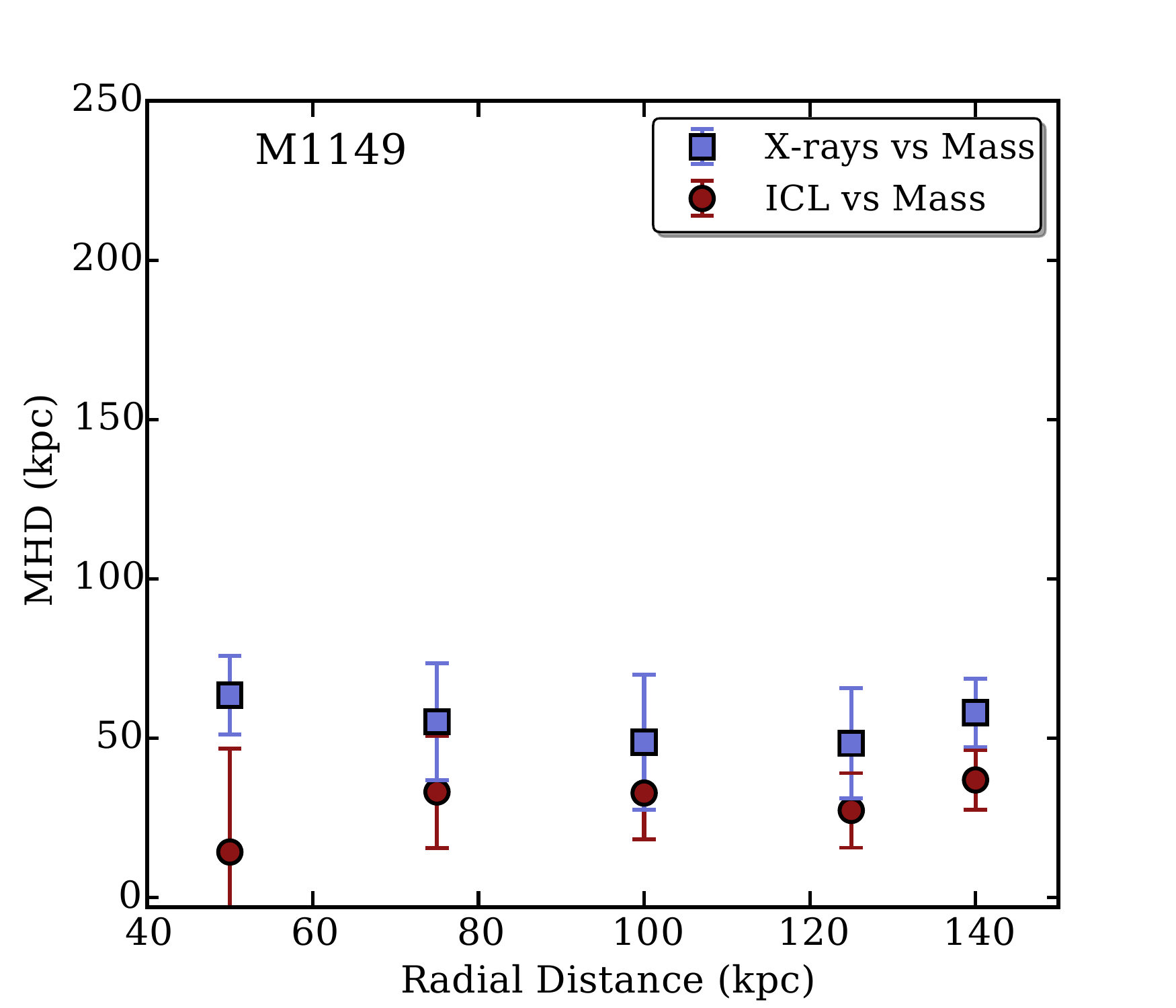}}
  \subfigure{\includegraphics[width=0.45\textwidth]{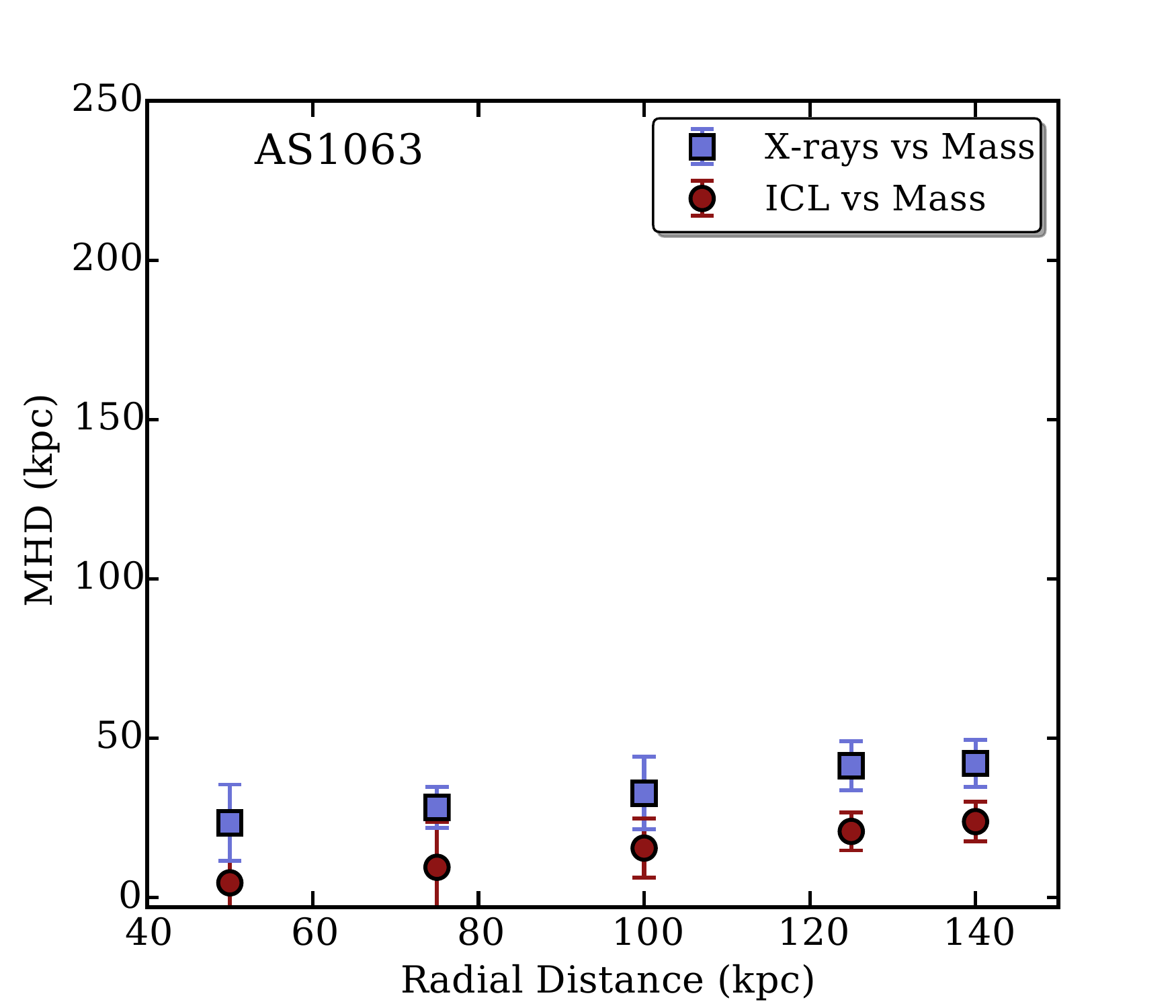}}
  \subfigure{\includegraphics[width=0.45\textwidth]{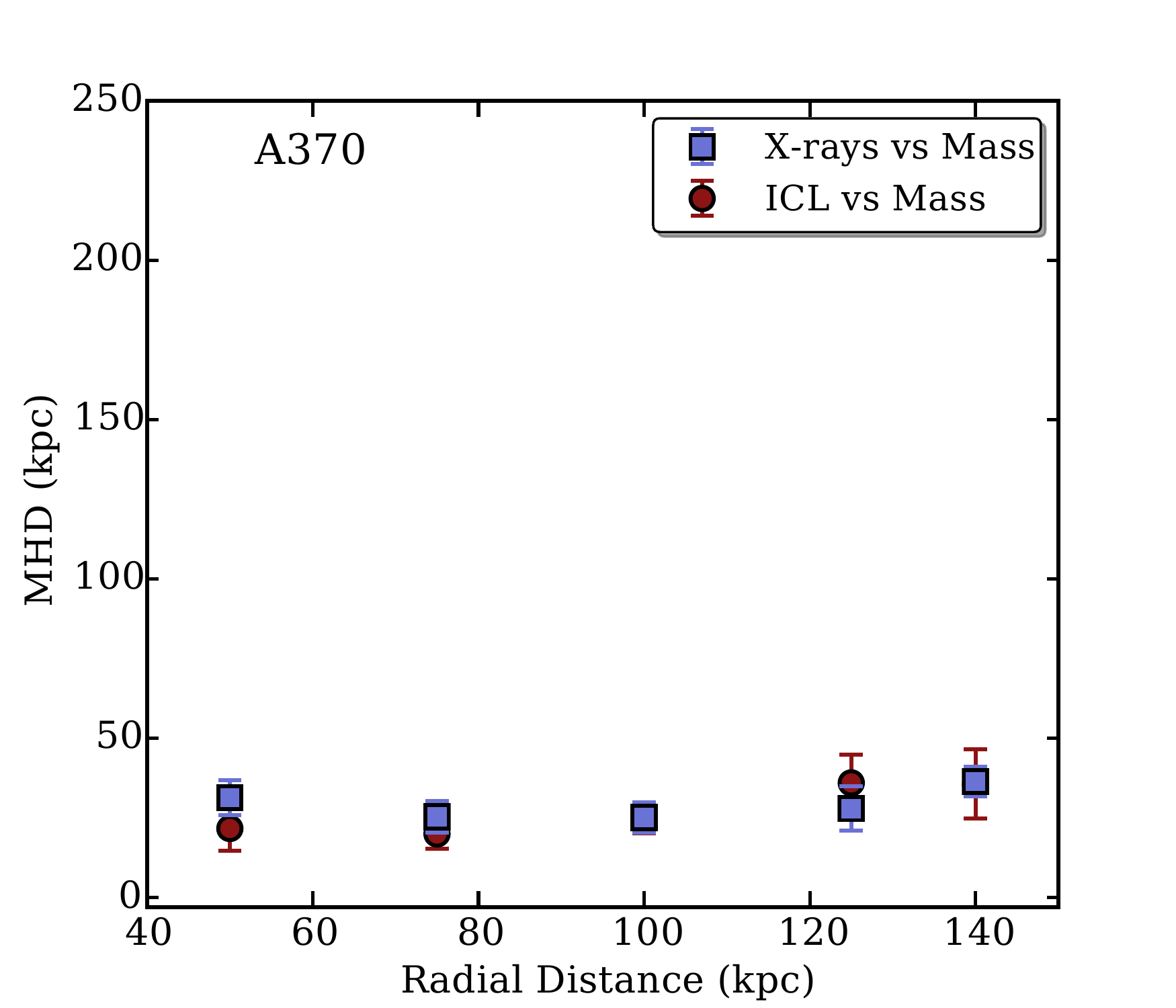}}
  
  \caption{Comparison between the MHD obtained by comparing the distribution of ICL with the mass models (red filled circles) and that comparing the X-rays distribution with the mass models (blue filled squares). In most of the cases, except A370, the X-rays distribution as compared with the underlying distribution of total mass has a large MHD, thus proving the power of the ICL as a tracer of dark matter halos.}\label{fig:mhdx}
\end{figure*}

\paragraph*{A2744} A2744 is the closest of the HFF clusters at $z=0.308$. It shows a significant degree of substructure \citep[][]{Jauzac2016b} and it is highly disturbed \citep[][]{Owers2011, Merten2011} undergoing a merger. Consequently, it is not surprising that the distribution of X-rays does not follow the mass distribution. However, we find a nice agreement between the ICL and the mass models. All this is quantified in the upper left panel of Fig. \ref{fig:mhd} and Fig. \ref{fig:mhdx}. Fig. \ref{fig:mhd} shows a median MHD between the ICL and the different mass model of $25$ kpc, compatible with the uncertainties among the different models (blue rectangles). Fig. \ref{fig:mhdx} shows that this agreement does not hold in the case of the distribution of X-rays. In fact, the average MHD between the X-ray distribution and the mass models is around $110$ kpc. This is consistent with \citet[][]{Owers2011} who found that there is not corresponding galaxy overdensity to the peak of the X-ray emission which interpreted as the stripped atmospheres of the merging subclusters. 

\paragraph*{M0416} M0416 is an elongated cluster at $z=0.397$ clearly undergoing a merging event as is shown by its X-ray morphology \citep{Mann2012}. Fig. \ref{fig:mhd} shows that the radial median value of the MHD between the ICL and the mass distribution is comparable to the uncertainties derived from the models, i.e. we find a good agreement between ICL and mass with a mean MHD of $17$ kpc. In the case of the X-rays, Fig. \ref{fig:mhdx} shows that while the difference is not as big as in the case of A2744, the X-ray distribution is still not as good tracer of the mass distribution of this cluster as the ICL. For the X-rays, the mean MHD is $28$ kpc.

\paragraph*{M0717} This cluster is the farthest \citep[$z = 0.545$,][]{Edge2003}, one of the most massive and the strongest lenser of all the clusters in the HFF sample \citep{Lotz2017}. This is a highly disturbed massive cluster \citep{Ebeling2007} with a complex structure and an ongoing merger \citep[][]{Ma2009} as shown by significant offsets between the X-ray emission and the mass distribution \citep[][]{vanWeeren2017}. In Fig. \ref{fig:mhd}, we find a good agreement between the ICL and the total mass distribution. However, contrary to what we can infer from Fig. \ref{fig:contourall} and Fig. \ref{fig:contourrgb}, the agreement at $>100$ kpc between X-rays and the distribution of mass in Fig. \ref{fig:mhdx} is not seen in the contours. This might be caused by the presence of a peculiarity that it is not seen in the other clusters (see Fig. \ref{fig:contourall}). In the inner $110\times 110$ arcsec$^2$ of this cluster, there are 3 different separate substructures. That creates three separate sets of contours for the ICL and the mass models. Moreover, the X-rays also present separate contours. For the future, it would be worth exploring whether the MHD might be able to properly deal with comparing complex shapes.

\paragraph*{M1149} A cluster at $z=0.543$, M1149 is an X-ray elongated cluster with a complex merger history \citep[][]{Kartaltepe2008, Zitrin2009, Lotz2017}. As mentioned in MT18, this cluster presents two bright stars and a foreground galaxy in the lower right part of the images. To prevent contamination on the ICL stellar populations, and to a lesser extent the shape of the ICL, we masked part of the image. That causes that in Fig. \ref{fig:contourall} only the half left part of the contours are shown. As we mentioned before, mass models in this cluster have a larger MHD due to the small number of multiple-imaged systems. This causes that the image positions predicted by the best-fitting models are less precise than in other clusters \citep[][]{Jauzac2016a}. The X-rays present an offset with respect to the ICL and the models, shown by the respective values of MHD in Fig. \ref{fig:mhdx}. The mean MHD for the ICL is $33$ kpc, while for X-rays is $60$ kpc. 

\paragraph*{AS1063} AS1063 is a cluster at $z=0.346$. Even though it is the most relaxed in the HFF sample \citep{Diego2016b}, it exhibits asymmetry in its X-ray emission map (Fig. \ref{fig:contourall}) suggesting a recent ($\sim 0.75$ Gyr) 1:4 mass ratio merger \citep[][]{Gomez2012}. The ICL traces the mass distribution of the cluster with a mean MHD of $16$ kpc as seen in Fig. \ref{fig:mhd}. The asymmetry of the X-rays with respect to the mass distribution and the ICL is shown in Fig. \ref{fig:mhdx}. As expected, the X-rays have a mean MHD larger than for the ICL: $34$ kpc.

\paragraph*{A370} A370 is one of the best studied strong-lensing clusters, $z=0.375$, as it is the host of the first observed Einstein ring  \citep[][]{Paczynski1987, Soucail1987}. \citet{Richard2010} found that this cluster is likely the recent merger of two subclusters of equal mass based on small offsets among X-rays, mass and stellar light (although no offset is present in \citealt{Lagattuta2017}). Fig. \ref{fig:contourall} shows the resemblance among the shape for the three tracers. The value for the mean MHD of the ICL is $24$ kpc (Fig. \ref{fig:mhd}). However, contrary to the other clusters, the X-rays in this cluster match very closely the 2D distribution of the ICL and, therefore, of the total mass of the cluster. The mean value for the MHD of the X-rays is $25$ kpc.

\section{Discussion}

The results presented in this work demonstrate the extraordinary power of the ICL to trace the shape of the total mass of the galaxy clusters. The depth, the multiwavelength coverage and the wealth of ancillary data available for the HFF clusters makes possible to have accurate total mass distributions from gravitational lensing, and therefore, to explore how well the ICL follows mass. 

\subsection{Is the X-ray emission a good tracer of the distribution of the total mass?}

The hot gas distribution, as traced by the X-ray emission maps, has been extensively used in the past as a luminous tracer for the total mass of the clusters \citep[e.g.][]{Borgani2001}. In this work, we have explored whether the ICL is a better and more detailed representation of the detailed distribution of the total mass in clusters than the hot gas. In most of the cases there is no similarity between the mass distribution and the X-ray emission. This is not surprising as the HFF clusters are all in the process or have experienced a recent merger. While in more relaxed clusters it might be true that the X-rays follow the DM distribution, in interactions, the gas (dissipative) experiences ram pressure and is slowed, creating an offset between the DM and the X-ray emission. However, one of the HFF clusters (although it is currently undergoing a merging process) does exhibit a good correspondence among ICL, X-rays and mass: A370. In \citet[][]{Richard2010}, the authors suggested that this agreement is the consequence of the two subcomponents of the cluster having a large projected velocity along the line of sight, which will explain the small offset seen between the X-rays and the DM peaks \citep[see also][]{Lagattuta2017}. 

Based on our quantitative analysis using the MHD, we conclude that the distribution of the ICL is a much better luminous tracer of the distribution of the total mass of the clusters than the X-ray emission. This is a significant step forward on our understanding of how the mass is located in these large scale structures. Having an accurate luminous signature for the total mass of the clusters will allow to make a detailed comparison with the dark matter distribution predicted by cosmological simulations.

\subsection{How accurate is the ICL as a tracer of the mass distribution of the cluster?}

\subsubsection{The accuracy of the mass models}
Despite the exquisite quality of the HFF data, the limited information given by the finite number of constraints (the multiple lensed images) produces a certain degeneracy in recovering the mass distribution maps. The intrinsic statistical errors of the HFF mass models (the rms between the predicted image position and the observation) are $\lesssim1.4\arcsec$ \citep[e.g.][]{Zitrin2015, Jauzac2015a, Lagattuta2017, Meneghetti2017, RemolinaGonzalez2018} which translates to distances of $\lesssim6$ ($\lesssim9$) kpc at redshift $z=0.3$ ($z=0.5$). This is less than the typical MHD that we found between the ICL (i.e. $\sim25$ kpc) and the mass models. However, these statistical errors of the models are not a good representation of the real uncertainties at recovering the underlying mass distribution. In fact, the multiple lensed images of a given cluster can be reproduced by different mass distributions hence a better description of the systematic uncertainties is provided by the dispersion of the mass models. We have characterized such uncertainty by the range of MHDs among the different mass models (blue rectangles in Fig. \ref{fig:mhd}). We found that the similarity between the total mass distribution and the ICL is within these systematic errors (see Fig. \ref{fig:mhd}). In this sense, with our current capabilities to characterize the mass distribution of the clusters, the ICL perfectly describes how the mass is distributed.


\subsubsection{Why the ICL traces the underlying mass distribution?}

In this paper we have shown that with the current level of precision at building mass models for galaxy clusters, the distribution of the ICL describes how the mass is distributed. A natural question then is why the ICL is such an accurate tracer of the underlying mass distribution? The answer to this is connected with the origin of the ICL. The ICL is the product of the stripping of galaxies as they fall into the cluster. Those stripped stars are not bound to any particular galaxy but to the cluster itself.  Furthermore, both the ICL and the DM act as collisionless components following the same gravitational potential. On the contrary, X-ray emitting gas is highly collisional, and consequently, its dynamic and spatial distribution does not follow the same rules as the collisionless components.


In order to explore how well the stellar halos trace their DM halos, \citet{Pillepich2018} analysed a sample of $4000$ galaxies in the IllustrisTNG simulation. Taking advantage of the large cosmological volume of the simulation (reaching $\sim300^3$ Mpc$^3$), they explored a wide range of halo masses from $\sim 1\times 10^{12}$ to $5\times10^{14}$ $M_{\odot}$. They found a correlation between the slope of the density profile of the stellar halo and the total mass of the system. These can be interpreted as a signature of the hierarchical assembly of halos in the $\Lambda$CDM paradigm. More massive halos form later which means that they are less concentrated \citep[e.g.][]{Navarro1996, Navarro1997, Gao2004}. These halos also tend to accrete more and more luminous satellites \citep[e.g.][]{Gao2004} at recent times. Those satellites tend to have larger apocentres and deposit their stars at large radii ($\sim100$ kpc, \citealt{Cooper2015}) forming a less centrally concentrated stellar profile.

The correlation between the slope of the density profile of the stellar halo and the total mass of the systems was already explored in MT18. We found that this relationship holds even at masses as large as the masses of the halos of the HFF clusters ($\sim3\times10^{15} M_{\odot}$). Furthermore, \citet{Pillepich2018} found that the stellar halo of the clusters could be as shallow as their dark matter halo. Inspired by this, in this paper we have shown that it is not only the global (1D) profile distribution of the ICL which agrees with the general slope of the dark matter halo but also its bi-dimensional structure.

The findings of this paper have an extraordinary consequence: the ICL is a powerful tool to study DM halos in large structures. Using this diffuse light has more advantages than other tracers of the gravitational potential of the cluster, e.g. X-rays. In fact, the observation of the ICL, although challenging, is less time consuming than other observables. For example, gravitational lensing requires not only deep images to identify the multiply-lensed images but also spectroscopy to confirm their redshifts. In comparison, for observing the ICL only deep imaging observations are required. Needless to say that, compared with X-ray data, the resolution achievable with optical data is better, allowing also to trace substructure at higher redshifts. On top of that, the HFF images target the centres of the clusters, where other phenomena (not necessarily related to the mass distribution) can contribute to the X-ray emission (e.g., active galactic nucleus).

\section{Conclusions}

The results presented in this work demonstrate the extraordinary power of the ICL to trace the detailed shape of the underlying dark matter halo of galaxy clusters. Taking advantage of the superb data and the mass models from gravitational lensing provided by the HFF Initiative, we quantified the similarity between the total mass of the clusters and the ICL. To do that, we adopted a metric used in object shape matching: the Modified Hausdorff distance. The MHD gives us the mean spatial difference (in kpc) between the bi-dimensional distributions of the total mass of the cluster and its ICL. We find that:

\begin{itemize}
\item The mean MHD difference between the total mass distribution and the ICL is  $\sim25$ kpc. This difference is similar to the typical difference among the different techniques of lens inversion to reconstruct the total mass maps. 

\item In most of the cases the X-rays are not tracing well the distribution of the total mass of the cluster as the hot gas is easily perturbed in merging clusters. The ICL is shown to be a much better tracer of the mass distribution than the hot gas.
\end{itemize}

In summary, the study of the spatial distribution of the ICL stands out as a promising way to infer, in high detail, the properties of the underlying DM halos in galaxy clusters. A great advantage of the ICL is that can be observable from ground-based telescopes too \citep[e.g.][]{Mihos2005, Krick2007, Iodice2017} where specially dedicated low surface brightness observations \citep[e.g.][]{Duc2015, Trujillo2016, Huang2018} will minimize observational biases. 

The next step will be to extend this analysis to larger scales to assess whether the similarity between the distributions of ICL and mass holds at larger cluster radius. In this sense, The Beyond Ultra-deep Frontier Fields And Legacy Observations (BUFFALO, GO:15117, PI: Steinhardt), an extension of the HFF survey, will provide a wider view of these clusters allowing us to explore the ICL extending out to the edges of the massive HFF clusters. Also, current and future facilities like HSC, LSST or WFIRST will provide wider field-of-views and statistics to study how the ICL traces the structures of groups and clusters in a wider range of total masses.

\section*{Acknowledgements}

We thank the anonymous referee for their comments that helped to improve this manuscript. We would also like to thank STScI directors M. Mountain, K. Sembach and J. Lotz, and all the HFF team for making these extraordinary data available. M.M thanks Sarah Brough and Priya Natarajan for their support and helpful discussions.
Support for this work was provided by NASA through grant HST-AR-14304 from the Space Telescope Science Institute, operated by AURA, Inc. under NASA contract NAS 5-26555 and by the Spanish Ministerio de Econom\'ia y Competitividad (MINECO; grant AYA2016-77237-C3-1-P). I.T. acknowledges financial support from the European Union's Horizon 2020 research and innovation programme under Marie Sklodowska-Curie grant agreement No 721463 to the SUNDIAL ITN network.
This work utilizes gravitational lensing models produced by PIs Brada\u c, Natarajan \& Kneib (CATS), Merten \& Zitrin, Sharon, Williams, Keeton, Bernstein and Diego, and the GLAFIC group. This lens modeling was partially funded by the HST Frontier Fields program conducted by STScI. STScI is operated by the Association of Universities for Research in Astronomy, Inc. under NASA contract NAS 5-26555. The lens models were obtained from the Mikulski Archive for Space Telescopes (MAST).
This work makes extensive use of the following software: Astropy \citep{Astropy2013, Astropy2018}, SExtractor \citep{Bertin1996}, NumPy \citep{Jones2001, Oliphant2007}, SciPy \citep{Jones2001,Oliphant2007}, Matplotlib \citep{Hunter2007}.




\bibliographystyle{mnras}
\bibliography{ff_bib.bib} 



\appendix

\section{Appendix: MHD values}
\vspace{30 pt}
We provide the median MHD values in kpc for the $5$ different radial distances from the centres of the mass/light distributions: 50, 75, 100, 125 and 140 kpc (see Sec. 3 for more details in the choice of the centre and derivation of the isocontours). Table \ref{table:mhd_icl} has the MHD values for the comparison among the ICL and the different mass models, Table \ref{table:mhd_models} for the comparison among the different mass models and Table \ref{table:mhd_xrays} for the comparison among the X-rays emission and the mass models. 

\begin{landscape}
\begin{table}
\begin{tabular}{ccccccc}
R (kpc) & A2744 & M0416 & M0717 & M1149 & AS1063 & A370 \\ \hline
50  & $8.1  \pm 12.3$ & $11.0 \pm 6.5$   & $35.3 \pm 19.4$ & $14.3 \pm 32.5$ & $4.6 \pm 21.2$ & $21.7 \pm 7.0 $\\
75  & $20.8 \pm 4.9$ & $11.4 \pm 6.0$    & $41.5 \pm 16.8$ & $33.1 \pm 17.6$ & $9.5 \pm 14.2$ & $20.0 \pm 4.7 $\\
100 & $25.0 \pm 5.0$ & $16.5 \pm 10.1$ & $35.8 \pm 6.2$  & $32.8 \pm 14.6$ & $15.5 \pm 9.3$ & $24.1 \pm 4.0$\\
125 & $31.0 \pm 4.8$ & $20.8 \pm 9.0$   & $41.2 \pm 10.8$ & $27.4 \pm 11.7$ & $20.8 \pm 6.0$ & $36.0 \pm 8.9$\\
140 & $32.9 \pm 5.6$ & $34.9 \pm 7.3$   & $41.0 \pm 14.3$ & $36.9 \pm 9.3 $ & $23.8 \pm 6.3$ & $35.6 \pm 10.9$\\
\end{tabular}\caption{Median MHD values and standard deviations of the comparison of the ICL with the different mass models. MHD is given in kpc.}
\label{table:mhd_icl}
\end{table}

\begin{table}
\begin{tabular}{ccccccc}
R (kpc) & A2744 & M0416 & M0717 & M1149 & AS1063 & A370 \\ \hline
50  & $9.8 \pm 13.3$   & $10.3 \pm 2.9$   & $62.1 \pm 32.2$ & $58.4 \pm 30.8$ & $7.9 \pm 2.4$  & $18.7 \pm 10.5$\\
75  & $21.3 \pm 10.1$ & $12.2 \pm 4.0$   & $45.7 \pm 20.7$ & $54.5 \pm 24.6$ & $18.6 \pm 18.7$ & $18.6 \pm 10.6$\\
100 & $29.5 \pm 10.2$ & $17.5 \pm 6.1$  & $39.9 \pm 12.3$ & $48.5 \pm 17.7$ & $31.7 \pm 15.8$ & $31.6 \pm 10.9$\\
125 & $29.1 \pm 11.1$ & $22.0 \pm 15.3$ & $41.5 \pm 19.9$ & $43.5 \pm 16.0$ & $32.0 \pm 13.0$ & $32.6 \pm 12.3$\\
175 & $32.1 \pm 10.8$ & $33.8 \pm 12.9$ & $52.2 \pm 23.7$ & $40.7 \pm 15.8$ & $26.3 \pm 9.7$  & $34.8 \pm 9.3$\\
\end{tabular}\caption{Median MHD values and standard deviations of the comparison among the different mass models. MHD is given in kpc.}
\label{table:mhd_models}
\end{table}

\begin{table}
\begin{tabular}{ccccccc}
R (kpc) & A2744 & M0416 & M0717 & M1149 & AS1063 & A370 \\ \hline
50  & $134.9 \pm 16.5$ & $27.3 \pm 9.0 $ & $231.3 \pm 28.3$ & $63.5 \pm 12.3$ & $23.5 \pm 12.0$ & $31.3 \pm 5.6$\\
75  & $93.2  \pm 12.6$ & $25.5 \pm 2.5 $  & $187.5 \pm 31.9$ & $55.2 \pm 18.3$ & $28.3 \pm 6.5$  & $25.3 \pm 5.0$\\
100 & $121.9 \pm 19.5$ & $32.8 \pm 7.8 $ & $56.0  \pm 8.1$  & $48.7 \pm 21.1$ & $32.8 \pm 11.4$ & $25.2 \pm 4.7$\\
125 & $116.9 \pm 18.5$ & $33.6 \pm 7.4 $ & $52.0  \pm 12.9$ & $48.4 \pm 17.2$ & $41.3 \pm 7.7$  & $28.0 \pm 6.9$\\
140 & $109.5 \pm 22.7$ & $40.7 \pm 5.0 $ & $43.1  \pm 19.3$ & $58.0 \pm 10.7$ & $42.1 \pm 7.3$  & $36.4 \pm 4.7$\\
\end{tabular}\caption{Median MHD values and standard deviations of the comparison of the X-rays and the mass model distributions. MHD is given in kpc.}
\label{table:mhd_xrays}
\end{table}
\end{landscape}


\bsp	
\label{lastpage}
\end{document}